\documentclass[11pt]{article}
\pdfoutput=1 

\usepackage{jheppub} 

\usepackage[T1]{fontenc} 

\usepackage{euscript,amsmath,amsthm,amsfonts,amssymb}
\usepackage[arrow,frame,matrix]{xy}
\usepackage{subcaption}

\newcommand{\R}{\mathbf{R}}

\newcommand{\D}{\mathcal{D}}
\newcommand{\F}{\mathcal{F}}

\DeclareMathOperator{\area}{area}
\DeclareMathOperator{\vol}{vol}

\title{\boldmath Riemannian and Lorentzian flow-cut theorems
}

\author[a]{Matthew Headrick}
\author[b]{and Veronika E.\ Hubeny}
\affiliation[a]{Martin Fisher School of Physics, Brandeis University, Waltham MA 02453, USA, and \\ Center for Theoretical Physics, Massachusetts Institute of Technology, Cambridge MA 02139, USA}
\affiliation[b]{Center for Quantum Mathematics and Physics (QMAP)\\
Department of Physics, University of California, Davis, CA 95616 USA}

\abstract{
We prove several geometric theorems using tools from the theory of convex optimization. In the Riemannian setting, we prove the max flow-min cut theorem for boundary regions, applied recently to develop a ``bit-thread'' interpretation of holographic entanglement entropies. We also prove various properties of the max flow and min cut, including respective nesting properties. In the Lorentzian setting, we prove the analogous min flow-max cut theorem, which states that the volume of a maximal slice equals the flux of a minimal flow, where a flow is defined as a divergenceless timelike vector field with norm at least 1. This theorem includes as a special case a continuum version of Dilworth's theorem from the theory of partially ordered sets. We include a brief review of the necessary tools from the theory of convex optimization, in particular Lagrangian duality and convex relaxation.
}

\preprint{BRX-TH-6325, MIT-CTP/4941}

\begin{document} 
\maketitle
\flushbottom

\section{Introduction}
\label{sec:intro}

The celebrated max flow-min cut (MFMC) theorem is a cornerstone of network theory, with applications in every field where networks appear. Given a network with a specified capacity on each edge, the theorem states that the maximum total flow from a source vertex to a sink vertex equals the minimum total capacity of the edges that need to be cut to separate the source from the sink. Both the maximum total flow and the minimum cut capacity are solutions of linear programs, and their equality is a special case of strong duality of linear programs.

MFMC admits a generalization in the setting of Riemannian geometry, which is fairly straightforward to state but not to prove \cite{Federer74,MR700642,MR1088184,MR2685608}. Here, a flow is defined as a divergenceless vector field with bounded norm, while a cut is a hypersurface in a given homology class. The theorem equates the maximum flux through the corresponding homology cycle to the minimum area of a cut. Both the maximum flux and the minimum area are solutions to convex (not linear) programs, and their equality can be proven using strong duality. As with any statement about minimal surfaces, a careful proof also requires the machinery of geometric measure theory.

The Riemannian MFMC theorem played a central role in the recent analysis by Freedman and Headrick of holographic entanglement entropy \cite{Ryu:2006bv,Ryu:2006ef} in terms of so-called ``bit threads'' \cite{Freedman:2016zud}. The Ryu-Takayanagi formula gives the entropy of a spatial region in a holographic field theory in terms of the area of the minimal surface homologous to the region in the dual bulk geometry. By the MFMC theorem, this area equals the maximum flux of a flow through the region. The integral curves, or flow lines, of the flow can be thought of as Planckian ``bit threads'' representing entanglement between different parts of the boundary. It was argued in \cite{Freedman:2016zud} that this picture resolves several conceptual issues raised by the Ryu-Takayanagi formula, as well as offering certain technical advantages compared to the usual minimal-surface formulation.

In the version of the MFMC theorem employed in \cite{Freedman:2016zud}, the roles of source and sink were played by regions on the boundary of a Riemannian manifold. To our knowledge, this particular version of the theorem has not actually appeared in the math literature. In Section \ref{sec:RMFMC} of this paper, we give a proof. Our proof is at a physicist's level of rigor; in particular, we do not address the issues of functional analysis and geometric measure theory required for a rigorous proof. We then prove various properties relating the max flow and min cut, for example that the max flow saturates the norm bound on the min cut and generically only there. Finally, we show that the so-called nesting properties for cuts and flows also follow from strong duality; these properties state that, for nested boundary regions, the cuts are nested and there exists a common max flow.

In Section \ref{sec:lorentzian}, we again apply convex optimization theory to prove a version of the MFMC theorem in the Lorentzian setting. Remarkably, the  fact that a Lorentzian spacetime does not have a positive-definite metric, and thereby does not admit a natural notion of minimal surfaces, does not preclude the formulation of a theorem  which is analogous to the Riemannian case. It turns out that this is possible thanks to the essential feature of a Lorentzian spacetime, namely  that it has precisely one time dimension, and thereby admits a causal structure.  
Because of this, MFMC now stands for \emph{min flow-max cut}, where a flow is a divergenceless timelike vector field with norm bounded \emph{below} pointwise, while a cut is an achronal hypersurface, or slice, homologous to a given boundary region. The Lorentzian MFMC theorem equates the \emph{minimum} flux of a flow through the corresponding homology cycle to the \emph{maximum} volume of a slice. Various choices of boundary conditions can be imposed on the slice. For example, in an asymptotically anti-de Sitter spacetime, one can anchor the slice to a given Cauchy slice of the timelike boundary, or let it float on the boundary. In the latter case, a boundary condition is imposed on the flow, namely that it must have vanishing normal component on the timelike boundary.

Several extensions and applications of this Lorentzian MFMC theorem are also given:
\begin{itemize}
\item The nesting property for flows and slices follows from the same reasoning as in the Riemannian case.
\item The theorem allows one to define the notion of a \emph{globally} maximal slice even in a setting where the slice volumes are infinite, such as asymptotically anti-de Sitter spacetimes without a cutoff.
\item For a causal spacetime and with certain boundary conditions, the theorem becomes a continuum version of Dilworth's theorem from the theory of partially-ordered sets.
\item A similar theorem holds for manifolds with metric of signature $(0,+,+,\ldots)$, such as null submanifolds of Lorentzian spacetimes. Thus, it provides a dual formulation for the maximal-area surface in a light sheet (including, for example, a holographic screen \cite{Bousso:1999cb}).
\end{itemize}
We mention two other possible applications of the theorem. First, it could be used as a starting point in developing numerical algorithms for finding maximal-volume slices. Second, it may be relevant to the study of holographic complexity, which in a certain version equates the volume of a maximal slice to the gate complexity of the state in the dual field theory \cite{Stanford:2014jda}.

We should emphasize that, although the Lorentzian MFMC theorem is most obviously applicable in the context of general relativity, our arguments are purely geometric---we do not restrict ourselves to spacetimes obeying the Einstein equation (or any other specified dynamics), nor do we impose any energy conditions or causality constraints.

As the theorems in this paper illustrate, the theory of convex optimization provides a powerful (and perhaps underappreciated) set of tools for problems in differential geometry, including many problems of importance to physics. To keep the paper relatively self-contained, we begin in Section \ref{sec:convex} with a brief exposition of the relevant parts of this theory. Further examples of geometric applications of convex optimization will be described in forthcoming work, including to a covariant formulation of bit threads in the context of holographic entanglement entropies \cite{covariantflows} and to minimal-area problems appearing in closed-string field theory \cite{closedsft}.

\section{Convex optimization}
\label{sec:convex}

Throughout this paper we will be using tools from the theory of convex optimization, including convex relaxation and Lagrangian duality. Since these tools may not be familiar to all physicists, we briefly review them in this section. For a highly readable introduction to convex optimization, including many examples, we recommend the textbook \cite{MR2061575}.

\subsection{Convex programs}\label{sec:basic}

We begin by setting up some necessary notation and terminology. A \emph{convex program} (or convex optimization problem) $P$ is defined by the following data:
\begin{itemize}
\item a vector space $Y$; we will denote the vectors by $y$;
\item a non-empty convex subset\footnote{
A \emph{convex} subset $\D$ is one that is closed under convex combinations, i.e.\ for any set of vectors $\{y_k\}\subset\D$ and 
numbers $\alpha_k\ge0$ such that $\sum_k\alpha_k=1$, we have $\sum_k\alpha_k y_k \in\D$.} $\D$ of $Y$ called the \emph{domain};
\item a convex function\footnote{
A function $f:\D\to\R$ is \emph{convex} if, for any set of vectors $y_k\in\D$ and numbers $\alpha_k\ge0$ such that $\sum_k\alpha_k=1$, $f(\sum_k\alpha_k y_k)\le\sum_k\alpha_k f(y_k)$. This is a weaker requirement than strict convexity, which would replace $\le$ by $<$ whenever more than one $\alpha_k$ is non-zero. For example, an affine function is convex but not strictly convex.} $f_0:\D\to\R$ called the \emph{objective} function;
\item a set of convex functions $f_i:\D\to\R$ called the \emph{inequality constraint} functions;
\item a set of affine functions\footnote{
A function $h:\D\to\R$ is \emph{affine} if, for any set of vectors $y_k\in\D$ and numbers $\alpha_k\ge0$ such that $\sum_k\alpha_k=1$, $h(\sum_k\alpha_k y_k)=\sum_k\alpha_k h(y_k)$. This is equivalent to $h$ being the sum of a linear and a constant function.} $h_j:\D\to\R$ called the \emph{equality constraint} functions.
\end{itemize}
A \emph{feasible point} $y$ is an element of $\D$ such that $f_i(y)\le0$ for all $i$ and $h_j(y)=0$ for all $j$; correspondingly, the \emph{feasible set} $\F$ of $P$ is the set of all feasible points,
\begin{equation}\label{feasible}
\F := \left\{y\in\D\,|\,f_i(y)\le0\ \forall i\,,\,h_j(y)=0\ \forall j\right\}.
\end{equation}
The conditions defining the domain $\D$ within $Y$ are called \emph{implicit constraints}, while the conditions $f_i(y)\le0$ and $h_j(y)=0$ defining the feasible set $\F$ within $\D$ are called \emph{explicit constraints}. The \emph{solution} $p^*$ of $P$ is the infimum of the objective $f_0$ on the feasible set:
\begin{equation}
p^* := \inf f_0(\F)\,.
\end{equation}
This infimum may be $-\infty$. On the other hand, the feasible set may be empty, in which case, since the infimum of the empty set is defined to be $+\infty$, we would have $p^*=+\infty$. An \emph{optimal point} $y^*$ is a feasible point at which $f_0(y^*)=p^*$. An optimal point may or may not exist, and if it exists it may or may not be unique. We will use the following notation to describe a convex program:
\begin{equation}\label{programdef}
P:\text{minimize }f_0(y)\text{ over }y\in\D\,,\text{ subject to }f_i(y)\le0 \ \forall i\text{ and }h_j(y)=0 \ \forall j,
\end{equation}
where the implicit and explicit constraints are placed before and after the words ``subject to'' respectively.

The definition of a given convex program contains some redundancies, and can be transformed in various ways to give a program that obviously has the same solution. Examples of such transformation include:
\begin{itemize}
\item writing explicit constraints as implicit ones and vice versa;
\item mapping $\D$ to a convex subset $\D'$ of $Y$ or another vector space by a bijective affine map, and pulling back the objective and constraint functions to $\D'$;
\item decomposing $Y$ into a direct sum $Y=Y_1\oplus Y_2$, so that any vector $y$ is written uniquely as $y=y_1+y_2$ ($y_1\in Y_1$, $y_2\in Y_2$), and finding the infimum of $f_0$ with respect to $y_1$ holding $y_2$ fixed, leaving a program in terms of $y_2$ (i.e.\ ``integrating out'' $y_1$);
\item conversely, adding extra variables with constraints in a way that, when integrated out, returns us to the original program.
\end{itemize}
As we will see, such transformations can be very useful. We will not attempt to define precisely the notion of equivalence of programs, but, roughly speaking, two programs are considered equivalent if it is obvious from their definitions---without actually finding their solutions---that those solutions agree.

At first glance, the distinction between implicit and explicit constraints may seem artifi\-cial---why not just work directly within the feasible set $\F$ from the start? However, defining the feasible set concretely in terms of inequality and equality constraint functions, rather than as an abstract convex set, allows the application of many powerful tools, both numerical and analytic. It is thus often advantageous to write as many constraints explicitly as possible. Similarly, it is often useful to add extra variables to a program, even when they may be trivially integrated out. We will see examples of this.

By an obvious generalization, we define a \emph{concave program} in terms of a concave objective function $g_0$, concave inequality constraint functions $g_i$, and affine equality constraint functions $h_j$:
\begin{equation}\label{concavedef}
P:\text{maximize }g_0(y)\text{ over }y\in\D\,,\text{ subject to }
g_i(y)\ge0 \ \forall i \text{ and }  \,h_j(y)=0 \ \forall j,
\end{equation}
The feasible set is the convex set
\begin{equation}
\F:=\{y\in\D\,|\,\forall i\,,\,g_i(y)\ge0\,;\,\forall j\,,\,h_j(y)=0\}
\end{equation}
and the solution is
\begin{equation}
p^*:=\sup g_0(\F)\,.
\end{equation}

Another tool we will use is \emph{convex relaxation}. This involves turning a non-convex optimization problem into a convex program by embedding the feasible set into a convex set and defining a convex objective function which is less than or equal to the original objective on the original feasible set. For example, an optimization problem on the integers could be replaced with one on the reals. A priori the solution to the convex program only gives a lower bound for the solution to the original program. However, in favorable cases, one can show that the solutions are the same; we will see several examples. Because convex programs are so much easier to solve and work with than general optimization problems, convex relaxation is often a very powerful tool.

\subsection{Lagrangian duality}\label{sec:duality}

Starting from a constrained optimization problem, Lagrangian duality involves introducing Lagrange multipliers to enforce the constraints and then solving for the original variables, thereby obtaining a new optimization problem for the Lagrange multipliers. Under certain conditions, the resulting ``dual'' problem is equivalent to the original ``primal'' one. A very similar procedure---introducing Lagrange multipliers and then integrating out the original variables---is sometimes used in physics, for example in the derivations of certain dualities of quantum field theories such as T-duality.

For physicists, Lagrange multipliers are most familiarly used to enforce equality constraints. The only new part with inequality constraints is that the corresponding Lagrange multipliers are themselves constrained to be non-negative. We will see explicitly how this works below, but it can be understood heuristically from the fact that the value of the Lagrange multiplier tells you how much ``force'' is required to enforce the constraint, and for an inequality constraint this force can have only one sign (consider a particle in a box; the walls of the box can only exert an inward force on the particle).

Given the convex problem $P$ defined in \eqref{programdef}, let $m$ and $n$ be the number of inequality and equality constraints respectively. We will use $\lambda$ to denote points in $(\R^+)^m$ (where $\R^+:=[0,\infty)$) and $\nu$ to denote points in $\R^n$. We now define the following function on $\D\times(\R^+)^m\times\R^n$:
\begin{equation}\label{Ldef}
L(y,\lambda,\nu) := f_0(y)+\sum_{i=1}^m\lambda_i\,f_i(y)+\sum_{j=1}^n\nu_j\,h_j(y)\,.
\end{equation}
In the convex optimization literature, $L$ is called the \emph{Lagrangian function}; we hope that this terminology will not cause too much confusion for our physicist readers. If we maximize $L(y,\lambda,\nu)$ with respect to $\lambda$ and $\nu$ for fixed $y\in\D$, we find that the supremum is finite if and only if $f_i(y)\le0$ for all $i$ and $h_j(y)=0$ for all $j$, in other words if and only if $y\in\F$. In that case the supremum equals $f_0(y)$. Thus we have
\begin{equation}\label{supL}
\sup_{\lambda\in(\R^+)^m\atop\nu\in\R^n}L(y,\lambda,\nu) = \begin{cases}f_0(y)\,,\quad &y\in\F\\+\infty\,,\quad &y\notin\F\end{cases}\,.
\end{equation}
Therefore,
\begin{equation}\label{Lagminmax}
\inf_{y\in\D}\sup_{\lambda\in(\R^+)^m\atop\nu\in\R^n}L(y,\lambda,\nu) =
\inf_{y\in\F}f_0(y)=p^*\,.
\end{equation}
Note that the infimum on the left-hand side is on $y$ in $\D$, not in $\F$, in other words we only need to impose the implicit, not the explicit, constraints, since the latter are already effectively accounted for in \eqref{supL}. As usual with Lagrange multipliers, we have traded constraints for extra variables. Moreover, since the Lagrange multipliers enforcing the inequality constraints are themselves constrained to be non-negative,  the number of inequality constraints is conserved.

The idea now is to switch the order of the minimization and maximization in \eqref{Lagminmax}. We thus define the dual domain $\D'$ as the subset of $(\R^+)^m\times\R^n$ on which $L(y,\lambda,\nu)$ has a finite infimum,
\begin{equation}\label{dualfeasible}
\D':=\{(\lambda,\nu)\in(\R^+)^m\times\R^n|\inf_{y\in\D}L(y,\lambda,\nu)>-\infty\}\,,
\end{equation}
and the dual objective $g_0:\D'\to\R$ as the value of that infimum:
\begin{equation}\label{g0def1}
g_0(\lambda,\nu):=\inf_{y\in\D}L(y,\lambda,\nu)\,.
\end{equation}
For fixed $\lambda\in(\R^+)^m$ and $\nu\in\R^n$, $L(y,\lambda,\nu)$ is a sum of convex functions of $y$ and is therefore itself a convex function of $y$. As can be verified by a short computation, this implies that $\D'$ is a convex set and $g_0$ is a concave function on it. Finding its maximum is thus a concave program, called the \emph{dual program} $P'$, where $P$ in this context is called the \emph{primal program}:
\begin{equation}\label{dualdef}
P':\text{maximize }g_0(\lambda,\nu)\text{ over }(\lambda,\nu)\in\D'\,.
\end{equation}
There are no explicit constraints, so the feasible set is just the dual domain, $\F'=\D'$. We'll call the solution $d^*$:
\begin{equation}\label{dsupg}
d^*:=\sup g_0(\D')\,.
\end{equation}

While the dual program \eqref{dualdef} does not have any explicit constraints, in practice $g_0$ often admits a natural extension to some larger convex subset of $(\R^+)^m\times\R^n$, with equality and/or inequality constraints such that the feasible set equals the right-hand side of \eqref{dualfeasible}. Such an equivalent program is often also referred to as the ``dual program''. A concave program can be dualized analogously; the details are given below (see the paragraph around \eqref{concaveLagrangian}). In many cases the dual of this dual is then the primal (or is equivalent to it).

Lagrangian duality is useful because of a combination of two facts. First, under quite general conditions, the solution of the dual equals that of the primal:
\begin{equation}
d^*=p^*\,.
\end{equation}
This rather non-trivial fact is called \emph{strong duality} (as opposed to \emph{weak duality}, which merely asserts that $d^*\le p^*$, and which always holds, as we will prove below). One simple condition that implies strong duality is \emph{Slater's condition}, which states that the primal problem admits a feasible (not necessarily optimal) point $y_0$ which is in the interior of the domain $\D$ and where all the inequality constraints are strictly satisfied, $f_i(y_0)<0$.\footnote{
Slater's condition is actually slightly weaker than what is described above, in two respects. First, $y_0$ only needs to be in the \emph{relative interior} of $\D$, which is the interior relative to the lowest-dimensional plane in $Y$ containing $\D$. Second, only inequality constraint functions $f_i$ that are non-affine need to be strictly negative at $y_0$; affine ones can satisfy $f_i(y_0)\le0$ as always for a feasible point.} A proof that Slater's condition implies strong duality, and much more discussion, can be found in any textbook on convex optimization, such as \cite{MR2061575}. At the end of this subsection, we will give a physicist's proof of strong duality.

Second, in many cases the objective and domain of the dual program can be expressed in closed form. As noted above, for fixed $\lambda\in(\R^+)^m$ and $\nu\in\R^n$, $L(y,\lambda,\nu)$ is a convex function of $y$ on $\D$. Therefore, to find $\D'$ and calculate $g_0$ still requires solving a convex program. The key, however, is that the explicit constraints of $P$ are not imposed in this program. Often those constraints are what makes the primal program hard to solve, and without them the infimum can be computed in closed form. This is one reason why one would choose to make constraints in the primal explicit rather than implicit, and even to add extra variables that can be trivially integrated out. When strong duality holds and $g_0$ can be computed in closed form, Lagrangian duality provides a concave program that is equivalent to the primal yet typically looks very different, and from which both analytical insights and numerical methods may be derived.

Two special cases of Lagrangian duality may be familiar to the reader. The first is duality of linear programs, in which $\D$ is the entire vector space $Y$ and the objective and all inequality constraint functions are affine. The second is the Legendre transform, in which $\D$ includes the origin of $Y$, there are no inequality constraints, and the only equality constraint is $y=0$; $g_0$ is then minus the Legendre transform of $f_0$.

Lagrangian duality is defined analogously in the case of a concave program \eqref{concavedef}. The Lagrangian is
\begin{equation}\label{concaveLagrangian}
L(y,\lambda,\nu) := g_0(y)+\sum_{i=1}^m\lambda_i\,g_i(y)+\sum_{j=1}^n\nu_j\,h_j(y)\,,
\end{equation}
which is defined on $\D\times(\R^+)^m\times\R^n$, the dual domain is
\begin{equation}\label{concavedualfeasible}
\D':=\{(\lambda,\nu)\in(\R^+)^m\times\R^n|\sup_{y\in\D}L(y,\lambda,\nu)<\infty\}\,,
\end{equation}
and the dual objective is
\begin{equation}
f_0(\lambda,\nu):=\sup_{y\in\D}L(y,\lambda,\nu)\,.
\end{equation}
This is a convex function, so the dual is a convex program. 
In this paper we will dualize both convex and concave programs, but for concreteness in the remainder of this section we will assume the primal is convex.

\subsubsection{Complementary slackness}\label{sec:complementary}

By the definitions \eqref{Ldef} and \eqref{g0def1}, for any $y\in\F$ and $(\lambda,\nu)\in\D'$,
\begin{eqnarray}\label{gap}
f_0(y)-g_0(\lambda,\nu)&\ge& f_0(y)-L(y,\lambda,\nu) \nonumber\\
&=& -\sum_i\lambda_if_i(y)-\sum_j\nu_jh_j(y)\nonumber\\
&\ge&0\,.
\end{eqnarray}
Thus, for feasible points, $f_0(y)\ge g_0(\lambda,\nu)$, which immediately implies weak duality:
\begin{equation}\label{weak}
p^*\ge d^*\,.
\end{equation}

Now let us assume that strong duality holds. Then $f_0(y^*) = g_0(\lambda^*,\nu^*)$ on any optimal points $y^*,\lambda^*,\nu^*$. In that case, the expression on the second line of \eqref{gap} must vanish. Since $y^*$ is optimal, $h_j(y^*)=0$, so the sum on $j$ vanishes. Furthermore, on any feasible point the terms in the sum on $i$ are non-negative; therefore they must vanish individually:
\begin{equation}\label{slackness}
\lambda^*_if_i(y^*)=0 \quad \text{(no sum).} 
\end{equation}
In other words, for each $i$, at least one of the two constraints $\lambda_i\ge0$ and $f_i(y)\le0$ is saturated. Equation \eqref{slackness} is called ``complementary slackness''.

Complementary slackness tells us that, for an inactive constraint, $\lambda^*_i$ vanishes. Even for an active constraint, the value of $\lambda_i^*$ can tell us quantitatively \emph{how} active the constraint is. Specifically, if the dual optimal point is unique, then tightening the $i$th constraint by $\epsilon$, i.e.\ requiring $f_i(y)\le-\epsilon$ rather than $f_i(y)\le0$, raises the solution (to first order in $\epsilon$) by $\epsilon\lambda_i^*$. This follows from the fact that the dual objective $g_0(\lambda,\nu)$ is shifted by $\epsilon\lambda_i$; if its maximizer is unique then the maximum is increased by $\epsilon\lambda_i^*$. Complementary slackness is a special case: tightening an inactive constraint has no effect on the solution, and indeed $\lambda_i^*=0$. Similarly, shifting an equality constraint by requiring $h_j(y)=-\epsilon$ raises the solution by $\epsilon \nu_j^*$. This additional information about the nature of the optimal point is another reason to make constraints explicit in the primal problem.

\subsubsection{Physicist's proof}

We will now give a heuristic explanation of strong duality based on a force-balance picture. Assume that the objective $f_0$ and constraints $f_i$, $h_j$ are differentiable, and that there exists an optimal point $y^*$ in the interior of the domain $\D$. At $y^*$, the force $F_0=-\nabla f_0(y^*)$ exerted by the objective must cancel the total force exerted by the constraints. An equality constraint function $h_j$ exerts a force normal to the plane $h_j(y)=0$ with an arbitrary coefficient $\tilde\nu_j$: $F_j=-\tilde\nu_j\nabla h_j$. Similarly, an active inequality constraint function $f_i$ (one for which $f_i(y^*)=0$) exerts a force normal to the plane $f_i(y)=0$ and directed into the allowed region $f_i(y)\le0$: $F_i=-\tilde\lambda_i\nabla f_i$, with $\tilde\lambda_i\ge0$. 
We thus obtain the force-balance condition
\begin{equation}
\nabla f_0(y^*) + \sum_{i \text{ active}}\tilde\lambda_i\, \nabla f_i(y^*) +\sum_j\tilde\nu_j\, \nabla h_j = 0\,, \qquad \tilde\lambda_i\ge0\ .
\end{equation}
We can include the inactive inequality constraints in this equation simply by setting $\tilde\lambda_i=0$ for them:
\begin{equation}\label{forcebalance}
\nabla f_0(y^*) + \sum_i\tilde\lambda_i\, \nabla f_i(y^*) +\sum_j\tilde\nu_j\, \nabla h_j = 0 \ ,
\end{equation}
\begin{equation}\label{kkt2}
\tilde\lambda_i\ge0\,,\qquad
\tilde\lambda_i\, f_i(y^*) = 0\,.
\end{equation}
(Note that for a given $y^*$ the $\tilde\lambda_i$ and $\tilde\nu_j$ satisfying \eqref{forcebalance}, \eqref{kkt2} are not necessarily unique.) We now observe that the left-hand side of \eqref{forcebalance} is the gradient with respect to $y$ of $L(y,\tilde\lambda,\tilde\nu)$. Since the latter is a convex function of $y$, any point where the gradient vanishes is a minimum, so by  the definition \eqref{g0def1} of $g_0$,
\begin{equation}\label{lastg}
g_0(\tilde\lambda,\tilde\nu) = L(y^*,\tilde\lambda,\tilde\nu)  \ . \end{equation}
By the definition \eqref{Ldef} of $L$, we have
\begin{equation}\label{lastL}
L(y^*,\tilde\lambda,\tilde\nu) = p^* +\sum_i\tilde\lambda_i\, f_i(y^*)+\sum_j\tilde\nu_j\, h_j(y^*) \ ,
\end{equation}
and by the second equation of \eqref{kkt2} along with the fact that on a feasible point $h_j(y)=0$, the RHS of \eqref{lastL} is just $p^*$. Combining \eqref{lastg} and \eqref{lastL}, we find that
\begin{equation}\label{last}
g_0(\tilde\lambda,\tilde\nu) = p^*\, .
\end{equation}
By the definition \eqref{dsupg} of $d^*$, \eqref{last} implies that $d^* \ge p^*$. Combined with weak duality, \eqref{weak}, this implies strong duality, with $(\tilde\lambda,\tilde\nu)$ being an optimal point of the dual program:
\begin{equation}
d^* = p^*\,,\qquad
(\tilde\lambda,\tilde\nu) = (\lambda^*,\nu^*)\,.
\end{equation}

The arguments of the previous paragraph can be formalized to prove that \eqref{forcebalance}, \eqref{kkt2} follow from the fact that $y^*$ is an optimal point. One can also prove conversely that \eqref{forcebalance}, \eqref{kkt2}, together with the feasibility conditions $f_i(y^*)\le0$, $h_j(y^*)=0$, imply that $y^*$ is optimal. Therefore, the problem of solving the primal is equivalent to the problem of solving this system of equations and inequalities, which are called the \emph{Karush-Kuhn-Tucker} (KKT) conditions.

\section{Riemannian max flow-min cut theorem}
\label{sec:RMFMC}

In this section we will prove a version of the max flow-min cut (MFMC) theorem on Riemannian manifolds, relating the maximum flux of a certain type of vector field to the minimum surface area in a given homology class. As we will see, a key step in the proof is the application of Lagrangian duality. Another step exemplifies the concept of convex relaxation, the replacement of a non-convex optimization problem by an equivalent convex program. The proof will thereby serve to illustrate the application of methods from the theory of convex optimization in the geometric context. It will also set the stage for the Lorentzian min flow-max cut theorem proved in the next section.

The Riemannian MFMC theorem is closely analogous to the well-known network MFMC theorem, which is described in many places (see e.g.\ \cite{wiki:mfmc}). The proofs are also closely analogous. To state and prove the Riemannian version in a mathematically careful way, however, involves subtle issues in functional analysis and geometric measure theory. In this paper, we will take a physicist's viewpoint and largely ignore these issues. Various versions of the theorem can be found in the mathematics literature (although not as far as we know precisely the one we prove here) \cite{Federer74,MR700642,MR1088184}; see also the nice overview in Chapter 2 of \cite{MR2685608}. The version we prove here concerns boundary regions of a general Riemannian manifold, and was applied in the paper \cite{Freedman:2016zud} in the context of holographic entanglement entropies.

After describing the set-up and stating the MFMC theorem in subsection \ref{sec:statement}, we prove it in subsection \ref{sec:proof} using Lagrangian duality. In subsection \ref{sec:relative}, we give a useful generalization involving homology relative to a boundary region. In subsection \ref{sec:relation}, we discuss the existence and uniqueness of max flows and min cuts and prove an interesting property relating them, which we then use to define the notion of a globally minimal surface in a non-compact manifold. Finally, in subsection \ref{sec:nesting} we generalize the previous analysis to nested boundary regions, showing that the min cuts and max flows satisfy respective nesting properties.

\subsection{Statement}\label{sec:statement}

In this section, $M$ is a compact oriented manifold-with-boundary equipped with a Riemannian metric $g_{\mu\nu}$.\footnote{
In the holographic context, $M$ would be a spatial slice of an asymptotically anti-de Sitter spacetime, with the asymptotic region excised along some cutoff surface. (If the conformal boundary is spatially non-compact, then those directions would be cut off as well.) The boundary of $M$ is the cut-off surface, possibly along with a black-hole horizon and/or an internal wall (e.g.\ orientifold or orbifold fixed plane or confining wall).} 
To establish convenient terminology, we define ``region'' and ``surface'' as follows. A \emph{bulk (boundary) region} is an embedded compact codimension-0 submanifold-with-boundary of $M$ ($\partial M$). A \emph{bulk surface} is an embedded compact oriented codimension-1 submanifold-with-boundary of $M$, whose interior is contained in the interior of $M$. We will keep track of the orientations of bulk surfaces and boundary regions using a unit normal covector field $n_\mu$. For the boundary $\partial r$ of a bulk region $r$, this covector is by convention outward-directed, so that Stokes' theorem takes the standard form
\begin{equation}\label{Stokes}
\int_r\sqrt g\,\nabla_\mu v^\mu = \int_{\partial r}\sqrt h\,n_\mu v^\mu\,,
\end{equation}
where $v^\mu$ is an arbitrary vector field and $h$ is the determinant of the induced metric on $\partial r$. In particular, for a boundary region, $n_\mu$ is always outward-directed.
\begin{figure}[tbp]
\centering
\includegraphics[width=0.4\textwidth]{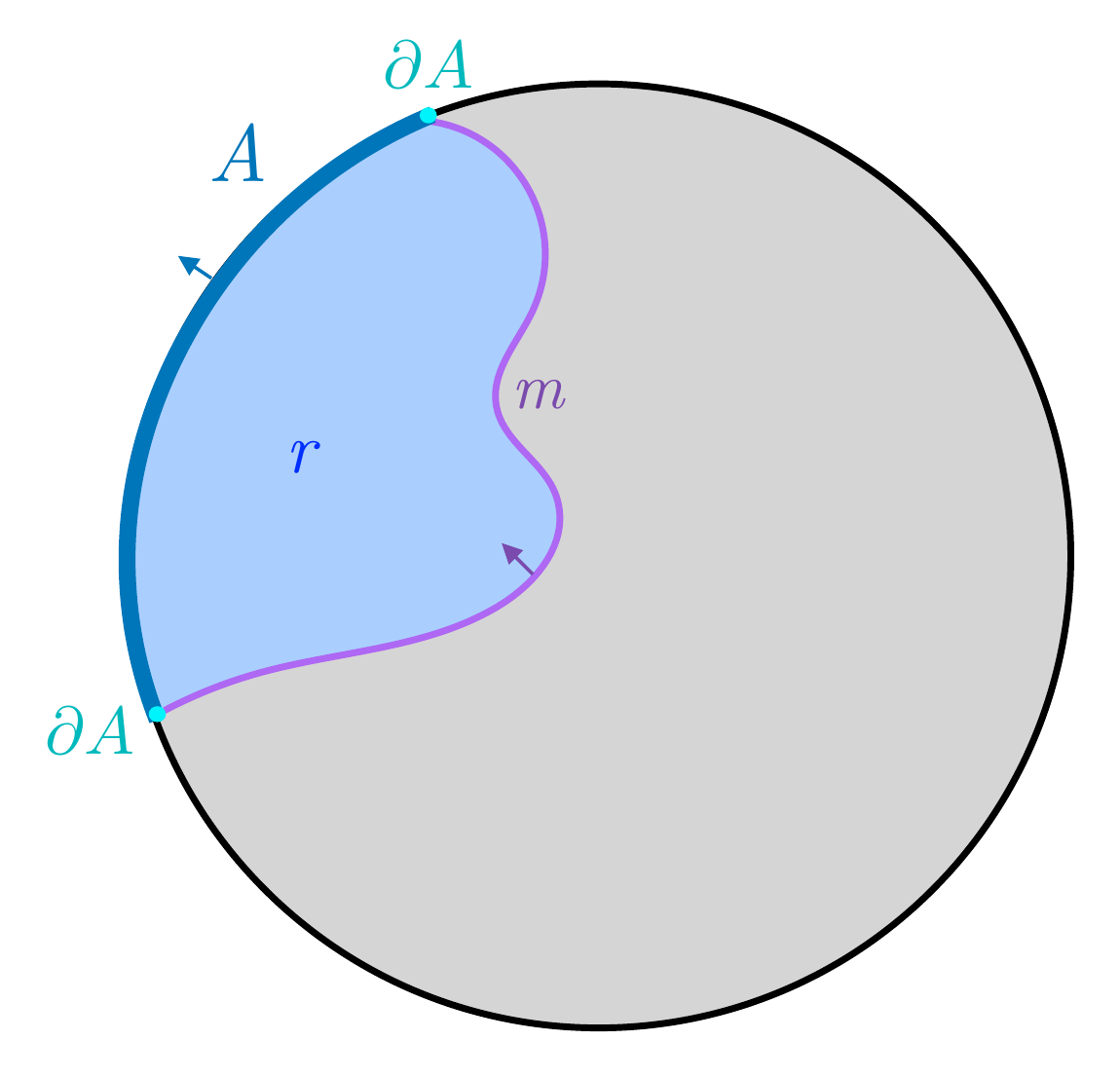}
\caption{\label{fig:homology}
Illustration of the homology condition.  The bulk surface $m$ is homologous to the boundary region $A$ ($m\sim A$) since there exists a region $r$ (shaded blue) with $\partial r=A-m$.  The normal covectors $n_\mu$ to $m$ and $A$ are indicated by the corresponding arrows.  (Note that the normal to $m$ is flipped compared to that of $\partial r$, which points outward from the region $r$.)
}
\end{figure}

Let $A$ be a boundary region. We say that a bulk surface $m$ is homologous to $A$ (writing $m\sim A$) if there exists a bulk region $r$ such that $\partial r=A-m$, where $A-m$ is the union of $A$ and $m$ with the orientation of $m$ flipped.\footnote{
More precisely, this is homology relative to $\partial A$. We will discuss homology relative to other subsets of $\partial M$ in subsection \ref{sec:relative}.} In particular, this requires $\partial m = \partial A$; see figure \ref{fig:homology}.

The two essential ingredients we consider in this paper are ``flows'' and ``cuts''.
We define a \emph{flow} as a vector field\footnote{
Via $w=*(v_\mu dx^\mu)$, flows $v^\mu$ are in one-to-one correspondence with $d-1$ calibrations $w$ on $d$-dimensional $M$.} $v^\mu$ on $M$ such that
\begin{equation}
\nabla_\mu v^\mu =0\,,\qquad |v^\mu|\le1\, ,
\end{equation}
and a \emph{cut} as any bulk surface $m\sim A$.
It follows from the divergencelessness condition and Stokes' theorem that, for any flow $v^\mu$ and any cut $m$,
\begin{equation}\label{mAflux}
\int_A\sqrt{h}\,n_\mu v^\mu =
\int_m\sqrt{h}\,n_\mu v^\mu \,.
\end{equation}
Furthermore, the condition $|v^\mu|\le1$ gives $n_\mu v^\mu\le1$, and therefore
\begin{equation}\label{mcondition}
\int_m\sqrt{h}\,n_\mu v^\mu \le\int_m\sqrt h= \area(m)\,.
\end{equation}
Combining \eqref{mAflux} and \eqref{mcondition}, we have
\begin{equation}\label{Acondition}
\int_A\sqrt{h}\,n_\mu v^\mu \le\area(m)\,.
\end{equation}
Thus, every surface homologous to $A$ serves as an obstruction to increasing the flux of $v^\mu$.

The MFMC theorem says that the respective supremum and infimum of the two sides of \eqref{Acondition} saturates the inequality:
\begin{equation}\label{mfmc}
\sup_{v^\mu}\int_A\sqrt{h}\,n_\mu v^\mu =\inf_{m\sim A}\area(m)\,.
\end{equation}
In other words, there are no obstructions other than \eqref{Acondition} to increasing the flux of $v^\mu$.

\subsection{Proof}\label{sec:proof}

The proof proceeds in three steps:
\begin{enumerate}
\item Show that the left-hand side of \eqref{mfmc} is the solution to a concave program satisfying Slater's condition.
\item Dualize the program to obtain an equivalent convex program.
\item Show that the solution to the convex program equals the right-hand side of \eqref{mfmc}.
\end{enumerate}

\paragraph{Step 1}The definition of a flow involves a linear equality constraint, $\nabla_\mu v^\mu=0$, and a concave inequality constraint, $1-|v^\mu|\ge0$. The functional $\int_A\sqrt h\, n_\mu v^\mu$ is linear. The left-hand side of \eqref{mfmc} is thus the solution\footnote{
We remind the reader that the \emph{solution} of a concave (convex) program is the supremum (infimum) of the objective subject to the constraints. A feasible point (in this case, a vector field configuration $v^\mu$) achieving the solution is called an \emph{optimal point}.} to the following concave program:
\begin{equation}\label{maxflow}
\text{\emph{max flow}}:\\
\text{maximize }\int_A\sqrt h\,n_\mu v^\mu\text{ over vector fields }v^\mu\text{ , subject to }\nabla_\mu v^\mu=0\,,\,1-|v^\mu|\ge0\,.
\end{equation}
We have chosen to make both constraints explicit. The same result is obtained by making the norm bound $|v^\mu|\le1$ implicit, but by keeping it explicit we will obtain more information about the nature of the solutions (see subsection \eqref{sec:relation}).

The vector field $v^\mu=0$ is feasible and strictly satisfies the inequality constraint. Therefore Slater's condition is satisfied.

\paragraph{Step 2}To dualize the \emph{max flow} program \eqref{maxflow}, we introduce a Lagrange multiplier $\psi$ for the divergenceless constraint, and another one $\phi$ for the norm bound. (Both are scalar functions on $M$, with $\phi$ constrained to be non-negative.) The Lagrangian function is
\begin{equation}
L[v^\mu,\psi,\phi]=\int_A\sqrt h\,n_\mu v^\mu+\int_M\sqrt g\left(-\psi\, \nabla_\mu v^\mu+\phi\, (1-|v^\mu|)\right).
\end{equation}
Our task now is to maximize $L$ with respect to $v^\mu$. First, we get the derivative off of $v^\mu$ by integrating by parts:
\begin{equation}\label{Lag2}
L[v^\mu,\psi,\phi]=\int_{\partial M}\sqrt h\,n_\mu v^\mu\, (\chi_A-\psi)+\int_M\sqrt g\left(v^\mu\, \partial_\mu\psi-|v^\mu|\phi+\phi\right)
\end{equation}
where $\chi_A$ is the characteristic function for $A$ on $\partial M$ (equal to 1 on $A$ and 0 on $A^c:=\partial M\setminus A$). We can now maximize $L$ pointwise. 
Recall that we are not imposing any restriction on $v^\mu$.
On $\partial M$, the integrand is bounded above if and only if $\psi=\chi_A$, and then vanishes. In the interior of $M$, the part of the  integrand involving $v^\mu$ is bounded above if and only if $\phi\ge|\partial_\mu\psi|$, and then vanishes. The dual objective is then just $\int_M\sqrt g\, \phi$, and correspondingly the dual program is
\begin{equation}\label{mincut1}
\text{\emph{min cut 1}}:\\
\text{minimize }\int_M\sqrt g\,\phi
\text{ over scalar fields $\psi,\phi$ with $\psi|_{\partial M}=\chi_A$, $\phi\ge|\partial_\mu\psi|$}\,.
\end{equation}
The reason for the name ``\emph{min cut 1}'' will become apparent below.

\paragraph{Step 3} First, we eliminate $\phi$ from the \emph{min cut 1} program \eqref{mincut1}. Given the simple form of the objective, the minimum is clearly achieved by setting $\phi$ at each point to its minimum allowed value, namely $|\partial_\mu\psi|$. We thus have\footnote{
We remark that the \emph{min cut 2} program can be dualized back to obtain the \emph{max flow} program \eqref{maxflow}. In order to take the derivative off $\psi$, it is necessary first to introduce a covector field $w_\mu$ and explicit constraint $w_\mu=\partial_\mu\psi$, and to write the objective as $\int_M\sqrt g|w_\mu|$. From there the dualization is straightforward; we leave the details as an exercise for the reader.}
\begin{equation}\label{mincut2}
\text{\emph{min cut 2}}:\\
\text{minimize }\int_M\sqrt g\,|\partial_\mu\psi|
\text{ over scalar fields $\psi$ with $\psi|_{\partial M}=\chi_A$}\,.
\end{equation}

We will now show that the solution to the \emph{min cut 2} program equals $\inf_{m\sim A}\area(m)$. We first give an outline of the argument. For any given function $\psi$ we consider its level sets $\{ \psi(x) = p \}$. The integral over $p$ of the area of the level sets equals the objective $\int_M\sqrt g\, |\partial_\mu\psi|$. Furthermore, the boundary condition implies that the level sets for values of $p$ between 0 and 1 are homologous to $A$. Therefore, the integrated area cannot be less than the minimal area in that homology class. On the other hand, for any surface $m\sim A$, $\psi$ can be chosen so that all of its level sets lie on $m$, in which case $\int_M\sqrt g\, |\partial_\mu\psi|=\area(m)$. Therefore, the two functionals have the same infimum.

\begin{figure}[tbp]
\centering
\includegraphics[width=0.52\textwidth]{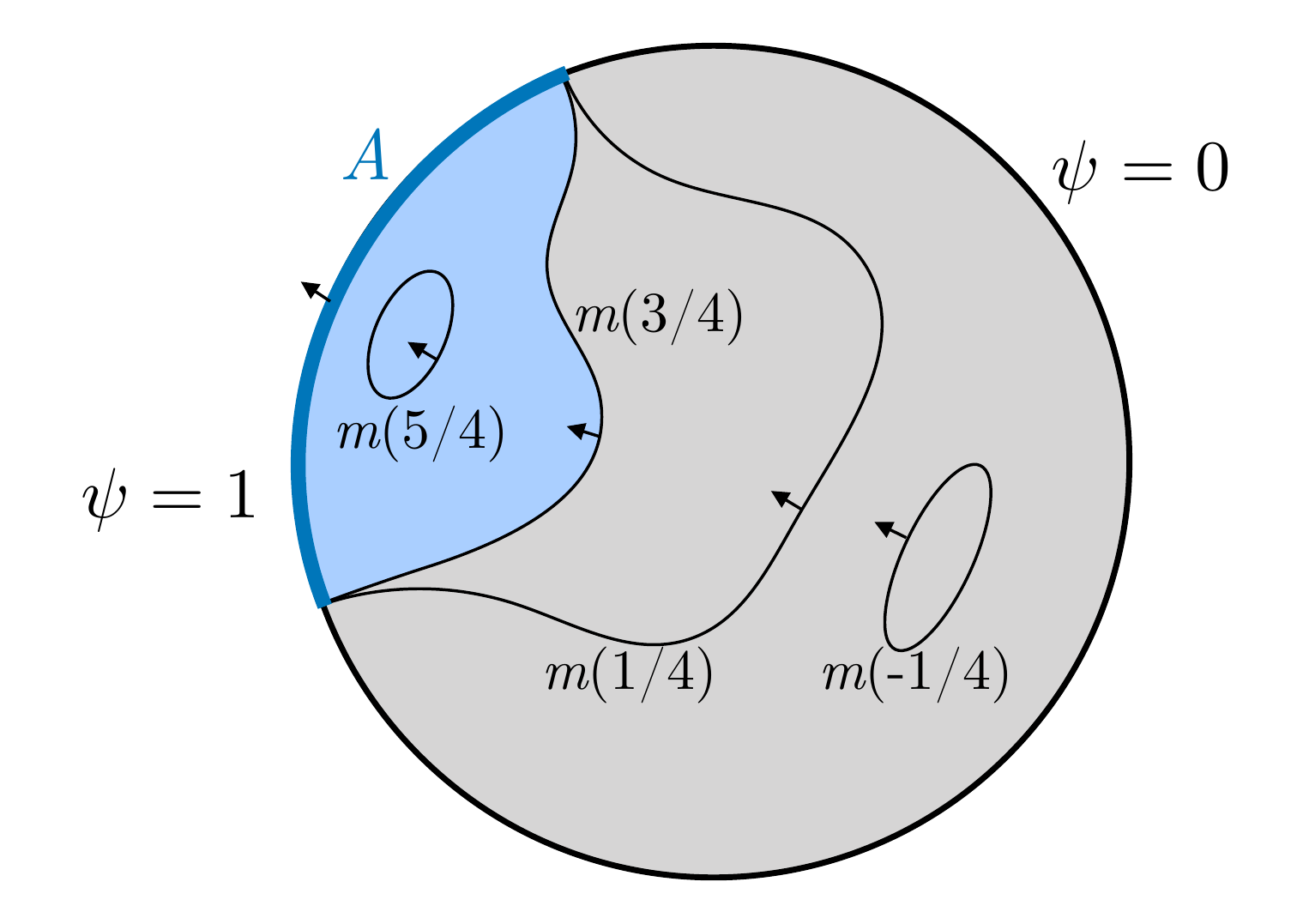}
\caption{\label{fig:levelsets}
Illustration of the level sets $m(p)$ defined below \eqref{rdef}. The boundary region $A$ is shown in dark blue. The boundary condition for $\psi$ sets it equal to $1$ on $A$ and $0$ on $A^c$. A selection of level sets is shown. The small arrows indicate their orientations. The bulk region $r(p)$ for $p=3/4$ is shown in light blue. As explained below \eqref{rBC}, the boundary condition on $\psi$ implies that $m(p)\sim A$ for $0<p<1$. As the figure illustrates, for $p>1$, $m(p)\sim\emptyset$, while for $p<0$, $m(p)\sim\partial M$.
}
\end{figure}

We now spell out this argument in more detail. We will assume for convenience that $\psi$ is differentiable (except on $\partial A$, where the boundary condition $\psi|_{\partial M}=\chi_A$ forces it to jump).\footnote{
The objective itself involves the gradient. It is therefore defined on non-differentiable functions as the limit of its value on differentiable ones (which are dense in the space of all functions). Hence, restricting to differentiable functions does not change the value of the infimum. A proper treatment of the functional analysis here can be found in the mathematical papers cited above.} Given a function $\psi$ on $M$ obeying the boundary condition $\psi|_{\partial M} = \chi_A$, define the one-parameter family of bulk regions $r(p)$ ($p\in\R$) as follows:
\begin{equation}\label{rdef}
r(p):=\{x\in M:\psi(x)\ge p\}\,.
\end{equation}
By the continuity of $\psi$, $\psi=p$ on $\partial r(p)\setminus\partial M$. The bulk surface $m(p)$, which we call the level set, is defined as the closure of $\partial r(p)\setminus\partial M$ with orientation covector $n_\mu$ pointed into $r(p)$ (i.e.\ parallel to $\partial_\mu\psi$); see figure \ref{fig:levelsets}. The objective can now be written in terms of the level sets as follows:
\begin{equation}\label{areas}
\int_M\sqrt g\,|\partial_\mu\psi| = \int_{-\infty}^{\infty} dp\,\area(m(p))\,.
\end{equation}
Equation \eqref{areas} can be shown by using $\psi$ as a coordinate on the part of $M$ where $\partial_\mu\psi\neq0$.

The boundary condition on $\psi$ implies that, for $0<p<1$,
\begin{equation}\label{rBC}
r(p)\cap\partial M=A\,.
\end{equation}
Hence the boundary part of $\partial r(p)$ equals $A$, while the bulk part equals $m(p)$ (by the definition of $m(p)$). Taking into account the orientation of $m(p)$, $\partial r(p) = A-m(p)$, hence $m(p)\sim A$ (see figure \ref{fig:levelsets}). Using this fact, together with the non-negativity of areas, the integrated area of the level sets is bounded below by the right-hand side of \eqref{mfmc}:
\begin{equation}
\int_{-\infty}^{\infty}  dp \,\area(r(p)) \ge \int_0^1 dp\,\area(r(p)) \ge
\inf_{m\sim A}\area(m)\,.
\end{equation}
Using \eqref{areas} and taking the infimum on the left-hand side, we have
\begin{equation}\label{psibound}
\inf_{\psi: \atop\psi|_{\partial M}=\chi_A}\int_M\sqrt g\,|\partial_\mu\psi|\ge\inf_{m\sim A}\area(m)\,.
\end{equation}

It remains to show that the bound \eqref{psibound} is saturated. Consider any bulk surface $m$ homologous to $A$. There is a bulk region $r$ such that $\partial r = A-m$. Let $\chi_r$ be its characteristic function (equal to 1 on $r$ and 0 on $M\setminus r$). This is not a differentiable function, since it has a jump on $m$. However, by slightly smoothing out the step, it can be approximated arbitrarily well by a differentiable function $\psi$. The level sets of $\psi$ then lie arbitrarily close to $m$, so by \eqref{areas} the objective $\int\sqrt g|\partial_\mu\psi|$ is arbitrarily close to $\area(m)$. Hence the infimum over all $m$ equals the infimum over all $\psi$:
\begin{equation}
\inf_{m\sim A}\area(m) = \inf_{\psi: \atop \psi|_{\partial M}=\chi_A}\int_M\sqrt g\,|\partial_\mu\psi|\,.
\end{equation}
This establishes that the solution to the \emph{min cut 2} program equals the right-hand side of \eqref{mfmc}, and completes the proof.

Step 3 of the proof is an example of convex relaxation. As discussed in subsection \ref{sec:basic}, this means replacing a non-convex optimization problem (in this case, finding the minimal-area surface in a given homology class) with an equivalent convex program (\emph{min cut 2}). Typically, this involves defining a convex set which is made up of convex or linear combinations of elements of the set on which the original problem is defined. Here, a general function $\psi$ obeying the boundary condition $\psi|_{\partial M}=\chi_A$ can be thought of as a linear combination of step functions $\chi_r$ (with $\partial r=A-m$); in this sense $\psi$ represents a ``smeared out'' surface. This is similar to passing from integral to real homology, although here, instead of linear combinations of homology classes, we are taking linear combinations of representatives of a single class.

\subsection{Relative homology}\label{sec:relative}

\begin{figure}[tbp]
\centering
\includegraphics[width=0.35\textwidth]{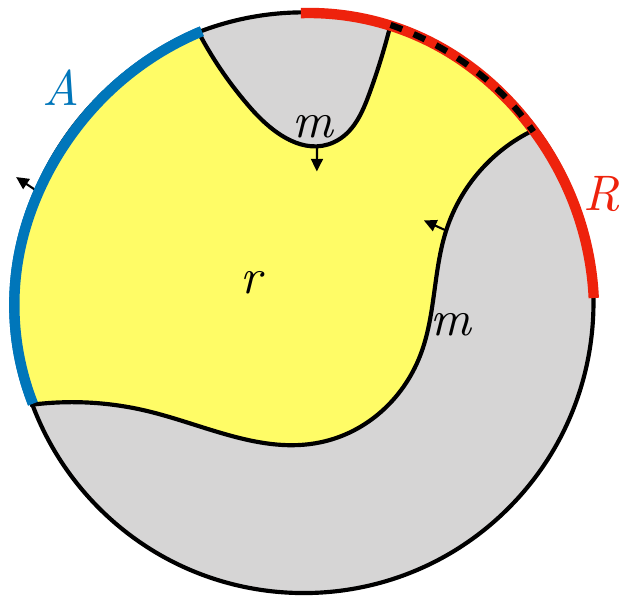}
\caption{\label{fig:relative}
Illustration of relative homology, defined in \eqref{relhomdef}. The boundary regions $A$ and $R$ are shown in blue and red, respectively. The bulk region $r$ is shown in yellow. The bulk surface $m$ obeys $m\sim A\text{ rel }R$ because \eqref{relhomdef} is satisfied. The left-hand side of that relation is shown as a dashed curve. (Although it is not shown in this figure, $R$ may also overlap $A$.)
}
\end{figure}

A useful generalization of the MFMC theorem involves loosening the boundary condition on the bulk surface $m$ while tightening the boundary condition on the flow $v^\mu$. Specifically, let $R$ be a boundary region.\footnote{
In the context of holographic entanglement entropy, $R$ could be an interior boundary of the bulk that does not carry entropy, such as a confining wall \cite{Klebanov:2007ws}, an orbifold or orientifold fixed plane, or the boundary $Q$ in AdS/BCFT duals \cite{Takayanagi:2011zk}. We'll see that according to \eqref{mfmc2}, the flow then has a Neumann (no-flux) boundary condition on $R$; in the language of \cite{Freedman:2016zud}, the ``bit threads'' are not allowed to end on $R$.} 
 We say that $m$ is homologous to $A$ \emph{relative to} $R$ (writing $m\sim A\text{ rel } R$) if there exists a bulk region $r$ such that $\partial r=A-m$ except possibly on $R$, in other words
\begin{equation}\label{relhomdef}
\partial r\setminus \partial M=-(m\setminus \partial M) \,,\qquad r\cap R^c=A\cap R^c
\end{equation}
where $R^c:=\partial M\setminus R$ (see figure \ref{fig:relative}). In order for the flux through $m$ to equal the flux through $A$ as in \eqref{mAflux}, we need to impose a Neumann boundary condition $n_\mu v^\mu=0$ on $R$. The generalized MFMC theorem is thus
\begin{equation}\label{mfmc2}
\sup_{v^\mu: \atop n_\mu v^\mu|_{{}_R}
=0}\int_A\sqrt{h}\ n_\mu v^\mu =\inf_{m\sim A\atop\text{ rel }R}\area(m)\,.
\end{equation}

The proof of \eqref{mfmc2} requires only minor alterations from the proof of \eqref{mfmc}. In Step 1, we add the implicit constraint $n_\mu v^\mu|_R=0$:
\begin{eqnarray}\label{maxflowrh}
\text{\emph{max flow}}:
\text{maximize }\int_A\sqrt h\ n_\mu v^\mu&\text{ over vector fields }v^\mu\text{ with }n_\mu v^\mu|_R=0\,,\nonumber \\ &\text{subject to }\nabla_\mu v^\mu=0\,,\,1-|v^\mu|\ge0\,.
\end{eqnarray}
Due to that implicit constraint, in Step 2, the boundary integrand in the Lagrangian function \eqref{Lag2} automatically vanishes on $R$, so the boundary condition on $\psi$ is imposed only on $R^c$:
\begin{equation}\label{mincut1rh}
\text{\emph{min cut 1}}:\\
\text{minimize }\int_M\sqrt g\,\phi
\text{ over scalar fields $\psi,\phi$ with $\psi|_{R^c}=\chi_A$, $\phi\ge|\partial_\mu\psi|$}\,.
\end{equation}
This implies that, in Step 3, instead of \eqref{rBC}, we only have
\begin{equation}
r(p)\cap R^c = A\setminus R\,,
\end{equation}
in turn implying $m(p)\sim A\text{ rel }R$. From there, the proof is the same as before, except with $m\sim A$ replaced by $m\sim A\text{ rel }R$ everywhere.

\subsection{Relation between max flow and min cut}\label{sec:relation}

As discussed in subsection \ref{sec:complementary}, Lagrangian duality does more than establish the equivalence of two programs. It also provides useful information about the optimal points $y^*$ and $(\nu^*,\lambda^*)$ for those programs. Complementary slackness tells us that the Lagrange multiplier $\lambda_i^*$ for an inequality constraint $g_i(y)\ge0$ must vanish if the constraint is inactive, i.e.\ $\lambda_i^*=0$ if $g_i(y^*)>0$. When the dual optimal is unique, we get even more information: tightening the constraint by $\epsilon_i$, i.e.\ imposing $g_i(y)\ge\epsilon_i$, reduces the solution by $\epsilon_i\lambda_i^*$. As we will show in this subsection, this fact has an interesting implication for the min cut and max flow.

First, we comment on the issue of existence and uniqueness of optimal points of the \emph{max flow} and \emph{min cut} programs. While we will not attempt to prove it, we expect a max flow always to exist. On a one-dimensional connected manifold (i.e.\ an interval), with $A$ being one endpoint, the max flow is clearly unique. In higher dimensions, one can construct examples where it is unique (e.g.\ $M=[0,1]\times M'$, where $M'$ is a closed $d-1$ manifold and $A=0\times M'$), but generically it is highly non-unique (specifically, it has a functional amount of non-uniqueness).

\begin{figure}[tbp]
\centering
\includegraphics[width=0.25\textwidth]{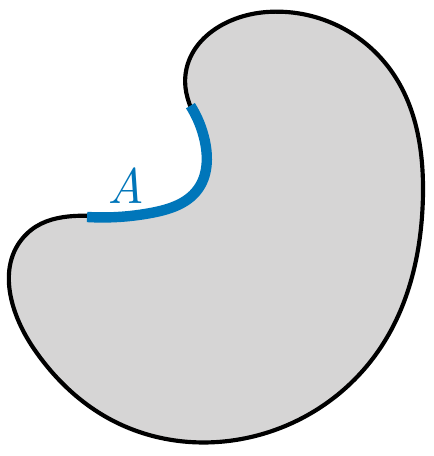}
\caption{\label{fig:notconvex}
A simple example where the minimal surface homologous to $A$ does not lie in the interior of $M$. Here $M$ is a region of the flat plane. The true minimal surface homologous to $A$ in this case is simply $A$ itself. Therefore, the infimum of the area over bulk surfaces is not achieved.
}
\end{figure}

For the min cut, the situation is slightly more complicated. We remind the reader that, in the statement of the theorem, we are taking the infimum over \emph{bulk} surfaces, defined as surfaces whose interior is contained in the interior of $M$. It may happen that bulk surfaces do not achieve the infimum, because the would-be minimal surface lies partially or entirely in $\partial M$ (see figure \ref{fig:notconvex} for an example).\footnote{
In the holographic context, for example, the minimal surface may wrap a horizon (which here we take as part of $\partial M$).} If we were to allow for such surfaces, then we would expect a min cut to exist, although again we will not attempt to prove it.\footnote{
Since a surface lying partially or entirely in $\partial M$ can be approximated arbitrarily closely by a bulk surface, allowing such surfaces does not change the value of the infimum. Interestingly, such surfaces automatically arise in the dualization if one makes the constraint $|v^\mu|\le1$ implicit rather than explicit as we did above: the dual program has no boundary condition on $\psi$ but rather a boundary term $\int_{R^c}\sqrt h|\psi-\chi_A|$ in the objective, which takes into account the area of the part of $m$ lying in $\partial M$.} In the examples above where the max flow is unique, the min cut is non-unique. However, in more than one dimension it is generically unique.

If there exist a min cut $m^*$ and max flow $v^{\mu*}$, then they must saturate \eqref{Acondition},
\begin{equation}
\int_A\sqrt h\,n_\mu v^{\mu*} = \area(m^*)\,.
\end{equation}
This implies that \eqref{mcondition} must be likewise saturated,
\begin{equation}
\int_{m^*}\sqrt h\,n_\mu v^{\mu*} = \area(m^*)\,,
\end{equation}
which in turn implies that $v^{\mu*}=n^\mu$ on $m^*$.
(In the bit thread language of \cite{Freedman:2016zud}, this means the threads are maximally packed on, and normal to, the minimal surface $m^*$.)

We now assume that a min cut $m^*$ exists and is unique. This is represented by the function $\chi_{r^*}$, where $r^*$ is the region bounded by $A-m^*$. In the \emph{min cut 1} program \eqref{mincut1rh}, the optimal value of $\phi$ is $\phi^*=|\partial_\mu\chi_{r^*}|=\delta_{m^*}$, a delta-function supported on $m^*$. Since $\phi$ is the Lagrange multiplier for the constraint $|v^\mu|\le1$ in the \emph{max flow} program \eqref{maxflowrh}, its value tells us how effective that constraint is. More precisely, $\phi^*$ is the decrease (to first order in $\epsilon$) of the maximum flux if we replace that constraint with $|v^\mu|\le 1-\epsilon$ (where $\epsilon$ is an arbitrary function on $M$). Since $\phi^*$ is supported on $m^*$, tightening the norm bound elsewhere has no effect (to first order) on the solution. Hence there exists a max flow where $|v^{\mu*}|<1$ everywhere except on $m^*$. In fact, for a generic choice of max flow $v^{\mu*}$, $|v^{\mu*}|=1$ \emph{only} on $m^*$.

As mentioned in \cite{Freedman:2016zud}, the fact that the min cut is the locus where $|v^{\mu*}|=1$ for every max flow allows us to define the notion of a globally minimal surface in a non-compact space, e.g.\ an asymptotically hyperbolic space with the surface anchored to a given surface in the conformal boundary.\footnote{
In such a space, we can define a {\emph locally} minimal surface as one where the trace of the extrinsic curvature vanishes; however, since their areas are typically infinite, given two such surfaces we cannot say which one has smaller area.
}  To explain this, we start on the flow side. A flow $v^\mu$ can be \emph{augmented} if there exists a vector field $\Delta v^\mu$ with positive flux through $A$ such that $v^\mu+\Delta v^\mu$ is still a flow. Clearly, in a compact space a flow that cannot be augmented is a max flow and vice versa. We can thus use this criterion as the \emph{definition} of a max flow, the advantage being that it can be carried over to the non-compact case, where the fluxes may be infinite. In turn, we can define a globally minimal surface in a non-compact space as the locus where all max flows have unit norm. We can compare this surface to the one obtained by first cutting off the space, then finding the globally minimal surface, and finally taking the limit where the cutoff is removed. In the presence of the cutoff the globally minimal surface is the locus where $|v^{\mu*}|=1$, so as long as the max flow changes continuously as the cutoff is removed, the two surfaces will agree.

\subsection{Nesting}\label{sec:nesting}

In this subsection we will further illustrate the power of strong duality by using it to establish two useful lemmas concerning the min cuts and max flows for nested boundary regions.

\subsubsection{Statement}

Before stating the lemmas, we set up some notation. Given a boundary region $A$, we will denote the maximal flux and minimal surface area, which are equal by the MFMC theorem, by $S(A)$.\footnote{
This notation originates from the holographic context, where this quantity equals an entanglement entropy.} We will also simplify the notation, dropping the index on $v^\mu$ and denoting its flux through $A$ simply by $\int_Av$, leaving the $\sqrt h\,n_\mu$ factors implicit. We also leave implicit the relative-homology region $R$; thus, $m\sim A$ means $m\sim A\text{ rel }R$ and all flows obey the Neumann boundary condition $n_\mu v^\mu|_R=0$. We thus have
\begin{equation}
S(A):=\sup_v\int_Av =\inf_{m\sim A} \area(m)\,.
\end{equation}
For simplicity, we will also assume that the infimum is uniquely achieved, i.e.\ there exists a unique minimal surface $m^*$. To make its dependence on the boundary region explicit we will denote it by $m(A)$, and the corresponding bulk region appearing in \eqref{relhomdef} by $r(A)$.

The lemmas concern the behavior of cuts and flows for nested regions on the boundary. In what follows, $A$ and $B$ denote arbitrary disjoint (but not necessarily separated, i.e.\ possibly sharing a common boundary) boundary regions, and $AB$ their union. The \emph{nesting property for flows} is the statement that there exists a flow $v(A,B)$ that simultaneously maximizes the flux through $A$ and through $AB$:
\begin{equation}
\int_{AB} v(A,B) = S(AB)\,,\qquad\int_Av(A,B) = S(A)\,.
\end{equation}
(Note that we do not require $v(A,B)$ also to maximize the flux through $B$, which may not be possible.) The \emph{nesting property for cuts} is the statement that the corresponding bulk regions are nested,
\begin{equation}\label{cutnesting}
r(A)\subset r(AB)\,;
\end{equation}
in other words, the function $r$ is monotonic with respect to inclusion.\footnote{
When the minimal surfaces are not unique, the precise statement is that they can be chosen to obey nesting. We will not consider this case here, but a proof can be found in \cite{Headrick:2013zda}.
} 

These two properties have been proven before using different methods. Nesting for cuts was proven in \cite{Headrick:2013zda} by a simple inclusion-exclusion argument.\footnote{
The argument is as follows: Define bulk regions $\tilde r(A):=r(A)\cap r(AB)$ and $\tilde r(AB):=r(A)\cup r(AB)$, and let $\tilde m(A)$, $\tilde m(AB)$ respectively be the bulk surfaces bounding them. If $r(A)\not\subset r(AB)$ then these are distinct from $m(A)$, $m(AB)$ and therefore have larger areas. However, by cutting and gluing the surfaces it is easy to see that $\area(\tilde m(A))+\area(\tilde m(AB))\le\area(m(A))+\area(m(AB))$, which is a contradiction.} 
Nesting for flows was proven in the network setting using the Ford-Fulkerson algorithm in \cite{Freedman:2016zud}.\footnote{
The argument is as follows: Starting from any max flow for $AB$, apply Ford-Fulkerson to find a max flow for $A$; this is possible by the greediness of Ford-Fulkerson. The only question is whether at the end we still have a max flow on AB. However, by definition the augmentations paths leave A, so they cannot reduce the flux on AB.} However, the proof below is the first one in the Riemannian setting that we are aware of. The proof given here is also interesting for the way that it links the two notions of nesting. In fact, we will show that strong duality simultaneously implies both properties.

\subsubsection{Applications}

These two properties have important implications in the context of holographic entanglement entropy. The nesting property for cuts is essential for the consistency of the so-called ``subregion duality'' conjecture, according to which the field-theory physics within the boundary region $A$ is represented holographically by the bulk region $r(A)$ \cite{Czech:2012bh,Wall:2012uf,Headrick:2013zda,Headrick:2014cta,Almheiri:2014lwa}. On the flow side, the proof of strong subadditivity requires nesting \cite{Freedman:2016zud}.

\begin{figure}[tbp]
\centering
\includegraphics[width=0.35\textwidth]{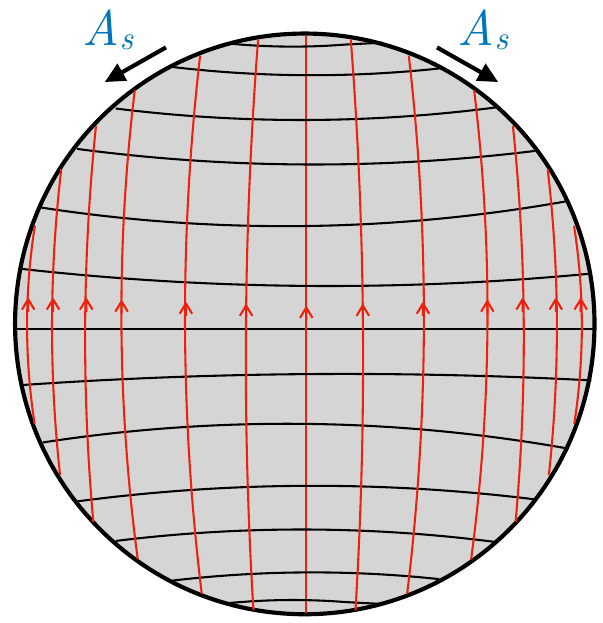}
\caption{\label{fig:foliation}
Example of nested minimal surfaces $m(A_s)$ and common max flow $v^*$ for a continuous one-parameter family of nested boundary regions $A_s$. The black curves are the minimal surfaces, and the red curves are the flow lines, or integral curves, of $v^*$. $A_0$ is the point at the top, and $A_1$ is all of $\partial M$. The boundaries $\partial A_s$ foliate $\partial M$, and the surfaces $m(A_s)$ foliate $M$.
}
\end{figure}

As noted previously, the max flow for a given region typically has a high degree of non-uniqueness. The nesting properties provide a simple way to fix some of this non-uniqueness, providing a class of canonically-defined flows. By a straightforward extension, they apply to multiple nested boundary regions $A$, $AB$, $ABC$, etc, and in fact to a continuous one-parameter family of nested regions. Suppose that we have a one-parameter family of regions $A_s$ such that $A_s\subset A_{s'}$ for $s\ge s'$, and whose boundaries $\partial A_s$ foliate the boundary, or some part of it. As illustrated in figure \ref{fig:foliation}, the corresponding minimal surfaces $m(A_s)$ foliate the part of the bulk that they pass through (which may not be the entire bulk, even when the entire boundary is foliated). By a generalization of the construction above, these are the level sets of the function $\psi^*$ that minimizes the functional $\int_M\sqrt g\,|\partial_\mu\psi|$ subject to the boundary condition $\psi=\chi$ on $R^c$, where $\chi$ is any boundary function whose level sets are the boundary surfaces $\partial A_s$; for example we could set $\chi$ equal to $s$ on $\partial A_s$. Recall that, on the minimal surface, the max flow equals the unit normal. The common max flow $v^*$ for all of the regions must equal the unit normal on all of the surfaces $m(A_s)$, and is therefore uniquely fixed in the foliated bulk region. (This vector field is indeed divergenceless: in general the unit normal vector field to a foliation obeys $\nabla_\mu n^\mu=K$, where $K$ is the trace of the extrinsic curvature of the leaves, and here $K$ vanishes by virtue of their minimality.) Since the minimal surfaces are the level sets of $\psi^*$, its gradient is orthogonal to them. Hence $v^*$ is (up to a sign) just the normalized gradient of the cut function $\psi^*$:
\begin{equation}
v^{\mu*} = -\frac{\partial^\mu\psi^*}{|\partial_\mu\psi^*|}\,.
\end{equation}
Hence foliating a given bulk region by minimal surfaces automatically constructs for us a divergenceless unit-norm vector field in that region, and this vector field can be extended to a flow on the whole bulk.

\subsubsection{Proof}

We begin the proof on the flow side. Consider the max flow program for the sum of the flux through $A$ and through $AB$:
\begin{equation}\label{combinedflow}
\text{\emph{combined max flow}}:\\
\text{maximize }\left(\int_Av+\int_{AB}v\right)\text{ over }v\text{ , subject to }\nabla_\mu v^\mu=0\,,\,1-|v^\mu|\ge0\,.
\end{equation}
The solution to \eqref{combinedflow} is clearly bounded above by the sum of the separate maxima of the two terms:
\begin{equation}\label{combinedflowbound}
\sup_v\left(\int_Av+\int_{AB}v\right)\le S(A)+S(AB)\,.
\end{equation}
Furthermore, if the bound \eqref{combinedflowbound} is saturated, then nesting for flows is obeyed.

\begin{figure}[tbp]
\centering
\includegraphics[width=0.45\textwidth]{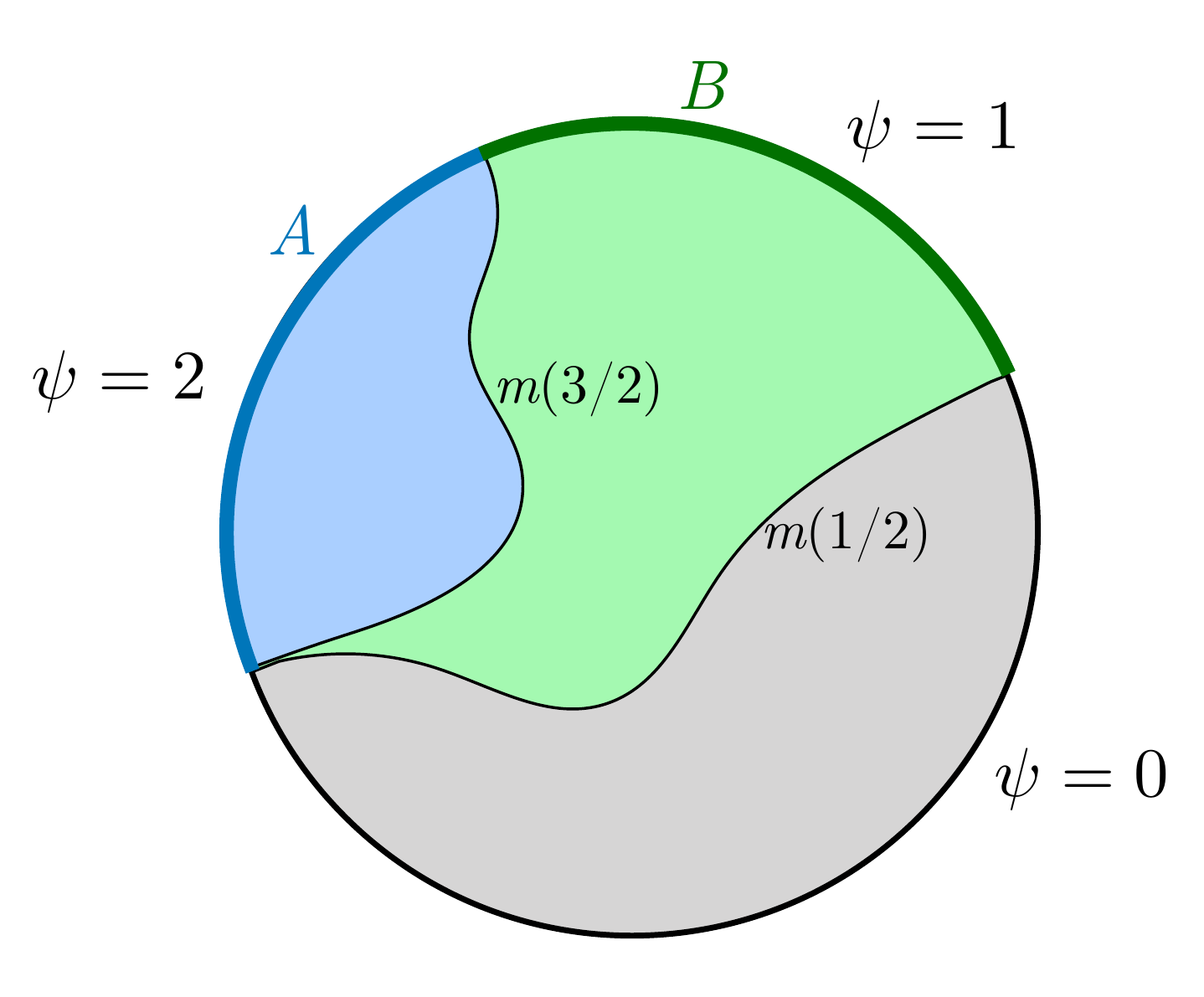}
\caption{\label{fig:nestedlevelsets}
Illustration of level sets $m(p)$ for $\psi$ obeying the boundary condition $\psi|_{R^c}=\chi_A+\chi_{AB}$ of the combined min cut program \eqref{combinedmincut}. (For clarity we set $R=\emptyset$.) For $0<p<1$, $m(p)\sim AB$, while for $1<p<2$, $m(p)\sim A$. The bulk region $r(3/2)$ is shown in blue, while the bulk region $r(1/2)$ is the union of the green and blue regions.
}
\end{figure}

The dual of \eqref{combinedflow} is (after eliminating $\phi$)
\begin{equation}\label{combinedmincut}
\text{\emph{combined min cut}}:\\
\text{minimize }\int_M\sqrt g\,|\partial_\mu\psi|
\text{ over $\psi$ with $\psi|_{R^c}=\chi_A+\chi_{AB}$}\,.
\end{equation}
Of course, $\chi_A+\chi_{AB}=2\chi_A+\chi_B$. The boundary condition on $\psi$ implies that its level sets $m(p)$ are homologous to $AB$ for $0<p<1$ and homologous to $A$ for $1<p<2$ (see figure \ref{fig:nestedlevelsets}). Therefore
\begin{equation}
\int_M\sqrt g\,|\partial_\mu\psi|=\int_{-\infty}^{\infty} dp\,\area(m(p))\ge\int_0^1dp\,\area(m(p))+\int_1^2dp\,\area(m(p))\ge S(AB)+S(A)\,.
\end{equation}
The only way this can be saturated is if all the level sets for $0<p<1$ fall on $m(AB)$, all the level sets for $1<p<2$ fall on $m(A)$, and the level sets for $p<0$ and $p>2$ vanish. Furthermore, by the definition \eqref{rdef} of $r(p)$, $r(p)\subset r(p')$ for $p>p'$, so saturation implies $r(A)\subset r(AB)$. To summarize,
\begin{equation}\label{combinedcutbound}
\inf_{\psi:\atop\psi=\chi_A+\chi_{AB}\text{ on }R^c}\int_M\sqrt g\,|\partial_\mu\psi| \ge S(A)+S(AB)\,,
\end{equation}
and if \eqref{combinedcutbound} is saturated then the nesting property for cuts holds.

The argument of the previous paragraph can be restated as follows. By using the level sets of a single function $\psi$ to represent simultaneously the $A$ cut and the $AB$ cut, we are requiring from the outset that the corresponding regions be nested. Then we are asking whether imposing this constraint increases the minimum of the sum of the areas. Without the nesting constraint, the minimum of the sum of the areas is just the sum of the minimal areas, $S(A)+S(AB)$. If adding the nesting constraint does not increase the total area, i.e.\ if \eqref{combinedcutbound} is saturated, then the nesting must have been automatically satisfied.

We now complete the proof. Strong duality says that the left-hand sides of \eqref{combinedflowbound} and \eqref{combinedcutbound} are equal. It follows that both inequalities must be saturated, and therefore both nesting properties hold.

\section{Lorentzian min flow-max cut theorem}\label{sec:lorentzian}

In this section, we will prove a Lorentzian analogue of the Riemannian max flow-min cut theorem of the last section. This theorem relates the \emph{maximum} volume of spacelike hypersurfaces (or slices) to the \emph{minimum} flux of timelike vector fields.\footnote{
Note that we do not expect there to exist an analogous theorem for timelike hypersurfaces in higher than two dimensions, since they can have neither maximal nor minimal volume; the volume of such a hypersurface can always be increased or decreased by adding short-wavelength undulations with either spacelike or timelike wave vector.} We therefore call this the \emph{min flow-max cut} theorem. 

This theorem may have applications in general relativity, where maximal-volume slices play an important role since the constraint equations simplify on them. 
In the context of holographic entanglement entropy, maximal volume slices were initially considered---but ruled out---in \cite{Hubeny:2007xt} as a step in covariantizing the Ryu-Takayanagi proposal. 
Recently, they reappeared in the conjectures 
relating the gate complexity in the field theory to the volume of a maximal slices \cite{Stanford:2014jda}. Since the gate complexity is defined as the \emph{smallest} number of gates required to construct the given state, it is natural that it should be related to the minimum value of a geometric quantity. Indeed, in analogy to the ``bit thread'' picture of holographic entanglement entropy \cite{Freedman:2016zud}, one can imagine a ``gate-line'' picture of holographic complexity.  While a more recent conjecture takes the form of  ``complexity = action'' \cite{Brown:2015bva},
maximal volume slice  also reappeared in
 \cite{MIyaji:2015mia} as the dual of quantum information metric (or fidelity susceptibility).

In the next subsection, we will describe the set-up and state the theorem, and in subsection \ref{sec:proof2} we will prove it. In subsection \ref{sec:extensions}, we will describe extensions of the theorem analogous to the Riemannian ones described in subsections \ref{sec:relation} and \ref{sec:nesting}. In subsection \ref{sec:dilworth}, we will explain how the theorem can be viewed as a continuous version of Dilworth's theorem from the theory of partially ordered sets. Finally, in subsection \ref{sec:degenerate}, we will give an extension to manifolds with metric of signature $(0,+,+,\ldots)$ such as null submanifolds of Lorentzian spacetimes.

\subsection{Statement}\label{sec:statement2}

The set-up, statement, and proof of the Lorentzian min flow-max cut theorem are, for the most part, closely analogous to those for the Riemannian max flow-min cut theorem. However, there are a few important changes, which we will highlight.

In this section, $M$ is a compact, oriented, and time-oriented Lorentzian manifold-with-boundary.\footnote{
We do not impose any equation of motion or curvature or causality conditions on $M$.} $M$ could be, for example, an asymptotically de Sitter, flat, or anti-de Sitter spacetime with a suitable cutoff. As in the previous section, a \emph{bulk (boundary) region} is an embedded compact codimension-0 submanifold-with-boundary of $M$ ($\partial M$). As in subsection \ref{sec:relative}, we fix  a boundary region (possibly empty) with respect to which relative homologies will be defined, and set $R^c:=\partial M\setminus R$. We put two conditions on $R^c$. First, both its causal future and its  causal past must cover $M$:
\begin{equation}\label{Rcond}
J^+(R^c) = J^-(R^c) = M\,.
\end{equation}
Second, $R^c$ must be covered by timelike and spacelike regions, i.e.\ must not include any null regions. This restriction is for technical reasons and could be removed with a slightly different formulation of the theorem.\footnote{
The issue is that, with our definition of a slice (given below), a null region of $R^c$ may not be deformable to any slice. Such a region may present an inverse bottleneck, lower bounding the flux of a flow, that is not visible to any slice. For example, one can construct an example in which the left-hand side of \eqref{mfmc3} is zero, but there are no slices homologous to $A$ so the right-hand side is $-\infty$. This issue can be addressed by allowing slices to coincide with the boundary and carefully treating the null case.}

We continue to keep track of the orientations of hypersurfaces using a normal covector $n_\mu$. Where the hypersurface is spacelike or timelike, $n_\mu$ is assumed to be normalized. The flux element of a vector field $v^\mu$ through the hypersurface is then $\sqrt hn_\mu v^\mu$. In the null case, even though the volume element $\sqrt h$ vanishes and $n_\mu$ has an undefined normalization, the flux element remains well-defined, and we will write it as $\sqrt hn_\mu v^\mu$ to simplify the notation. With this notation in hand, we still have Stokes' theorem in the standard form:
\begin{equation}
\int_r\sqrt g\,\nabla_\mu v^\mu = \int_{\partial r}\sqrt h\,n_\mu v^\mu\,.
\end{equation}

A \emph{slice} is an embedded compact oriented codimension-1 submanifold-with-boundary of $M$, whose interior is contained in the interior of $M$, which is piecewise spacelike or null, and whose orientation covector is future-directed (i.e.\ $n_0>0$). Slices will presently play the role played in the Riemannian case by bulk surfaces. However, there are certain changes relative to that case. First, to conform to the standard terminology in general relativity, we will use the term \emph{volume} rather than \emph{area}, writing $\vol(\Sigma):=\int_\Sigma\sqrt h$, where $h$ is the determinant of the induced metric. Second, we will be \emph{maximizing} this volume. (Minimal-volume slices do not exist, since the volume of a slice can always be decreased by adding wriggles in the time direction, just as maximal-area surfaces don't exist in the Riemannian case.) Third,  there do not always exist slices homologous to a given boundary region, as we will see below.

The definition of (relative) homology is unchanged from the previous section. Let $A$ be a boundary region. Since the part of $A$ in $R$ plays no role, we will assume without loss of generality that $A$ does not overlap $R$. Given a slice $\Sigma$, we write $\Sigma\sim A$ (we leave the ``rel $R$'' implicit in this section) if there exists a bulk region $r$ such that
\begin{equation}\label{relhomdef2}
\partial r\setminus \partial M=-(\Sigma\setminus \partial M)\,,\qquad r\cap R^c=A\,.
\end{equation}
The first equation, together with the condition on $\Sigma$ that its orientation be future-directed, implies that a future-directed timelike curve that intersects $\Sigma$ must enter $r$, so $\Sigma$ is achronal.

There is an important difference with the Riemannian case. Whereas, in the latter case, there always exist bulk surfaces homologous to any given boundary region, in the Lorentzian context, this is not true. 
A simple example would be a situation where the boundary of $A$ is not achronal, since in that case a slice anchored on $\partial A$ could not itself be achronal.
A necessary and sufficient condition for the existence of a slice homologous to $A$ is\footnote{
Although not needed in the rest of the paper, for completeness we indicate the proof here. \\
{\bf Necessity:} Assume there exists a slice $\Sigma\sim A$. The first equation in \eqref{relhomdef2}, together with the orientation of $\Sigma$, implies that $\partial r\setminus \partial M$ has a past-directed normal, hence $r$ lies to the future of $\Sigma$. Therefore a future-directed causal curve can never leave $r$. By the second equation in \eqref{relhomdef2}, a point in $A$ is necessarily in $r$, so any future-directed causal curve starting at such a point stays in $r$; if it intersects $R^c$, then by the same equation it does so in $A$. Equation \eqref{Acond} follows. \\ 
{\bf Sufficiency:} Let $n$ be a bulk region such that $n\cap R^c=A$. Set $r$ equal to the closure of $J^+(n)\setminus J^-(R^c\setminus A)$ and $\Sigma$ equal to minus the closure of $\partial r\setminus \partial M$. By construction, $\Sigma$ is a slice and the first equation of \eqref{relhomdef2} is satisfied. By \eqref{Acond}, $J^+(A)$ doesn't intersect $R^c\setminus A$, or equivalently $J^-(R^c\setminus A)$ doesn't intersect $A$, from which the second equation of \eqref{relhomdef2} follows.
}
\begin{equation}\label{Acond}
J^+(A) \cap R^c = A\,.
\end{equation}

\begin{figure}[tbp]
\centering
\includegraphics[width=\textwidth]{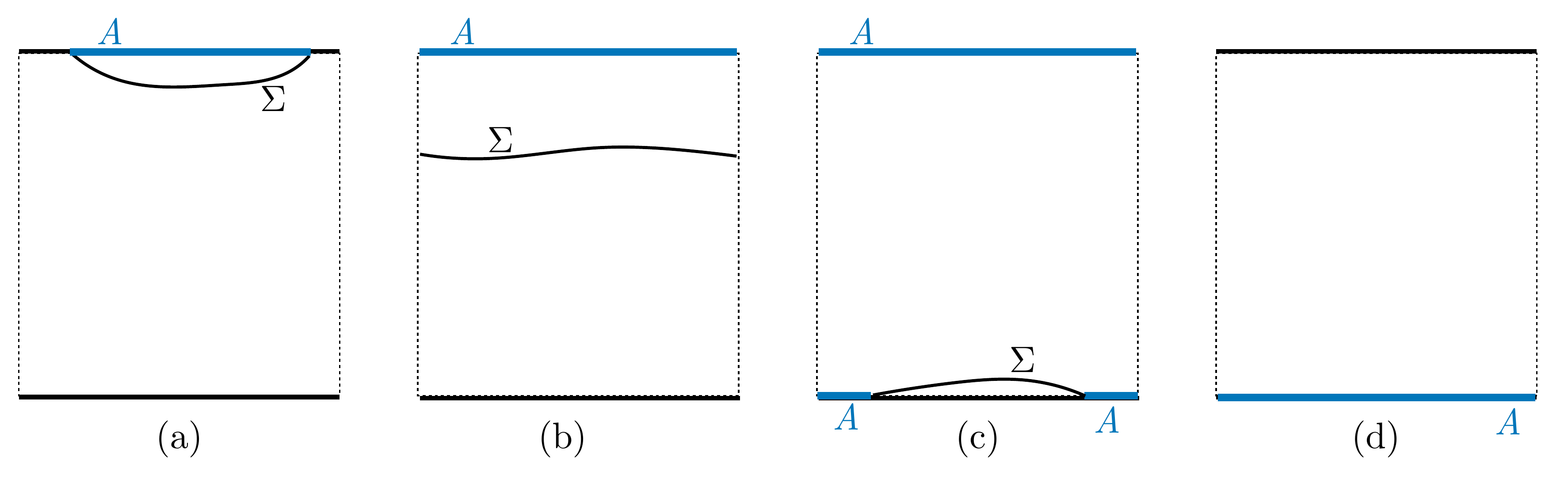}
\caption{\label{fig:deSitter}
Three choices of boundary region $A$ that satisfy \eqref{Acond} for an asymptotically de Sitter spacetime, and one that does not. The top and bottom boundaries of each rectangle are the cutoff surfaces near $\mathcal{I}^\pm$ respectively. (The left and right edges as usual represent the poles of spatial slices, not boundaries of the manifold.) (a) $A$ is a subset of the future cutoff; then any slice $\Sigma\sim A$ is anchored to the future boundary on $\partial A$. (b) $A$ covers the future boundary; $\Sigma$ is any Cauchy slice, and is not anchored to the boundary. (c) $A$ covers all of the future boundary and part of the past boundary; $\Sigma$ is anchored to the past boundary. (d) $A$ covers only the past boundary; there is no slice homologous to $A$, since any hypersurface homologous to $A$ has a past-directed normal.
}
\end{figure}

\begin{figure}[tbp]
\centering
\includegraphics[width=0.5\textwidth]{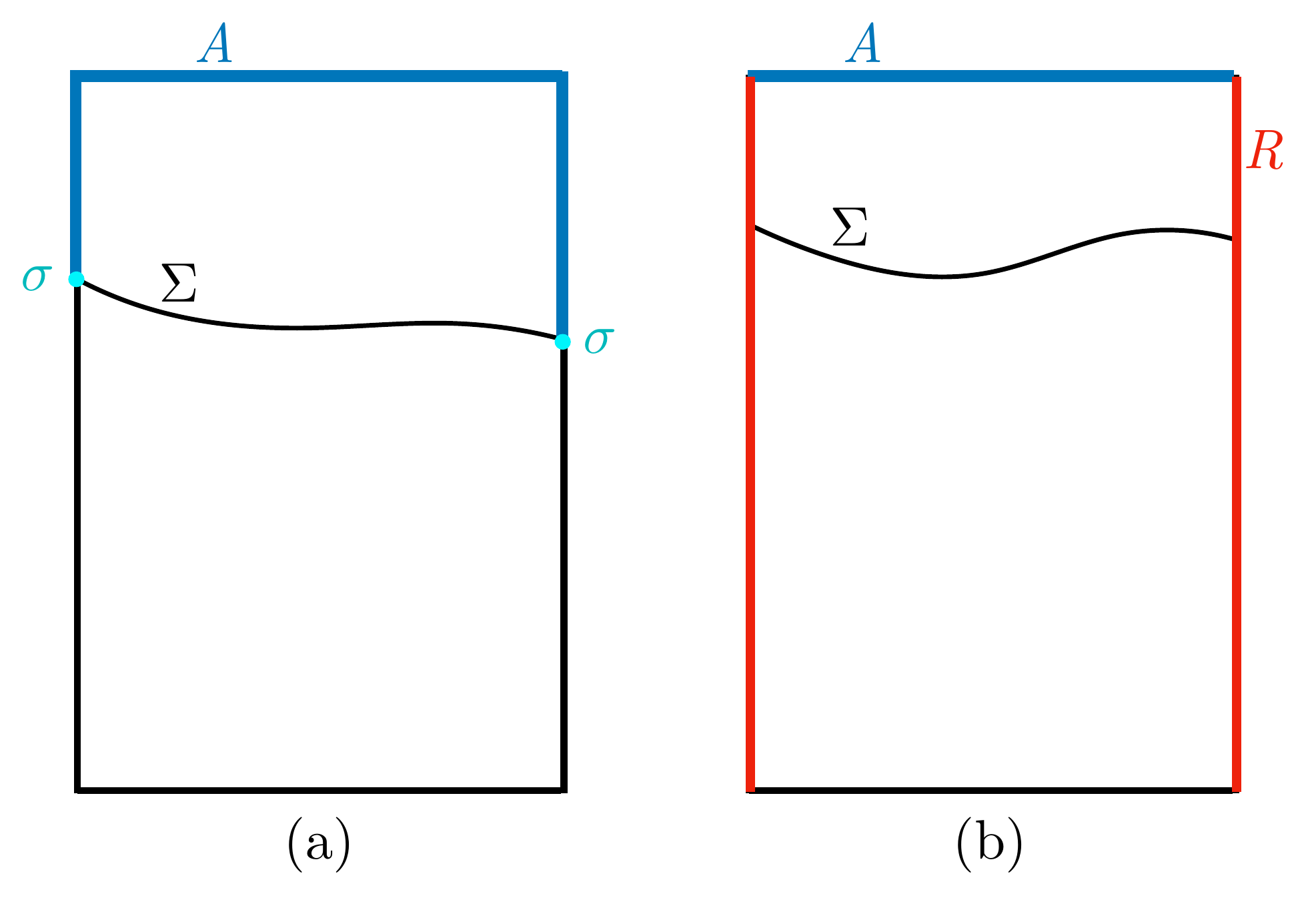}
\caption{\label{fig:AdS}
Two of the possible configurations of $A$ and $R$ satisfying \eqref{Acond} in an asymptotically AdS spacetime with spacelike and timelike cutoff boundaries. (a) $R=\emptyset$ and $A=J^+(\sigma)\cap\partial M$, where $\sigma$ is a Cauchy slice of the timelike boundary. Any slice $\Sigma\sim A$ is anchored to the timelike boundary along $\sigma$. A flow $v^\mu$ can have arbitrary flux through the timelike boundary. (b) $R$ is entire timelike boundary, and $A$ is the future boundary. Any Cauchy slice is homologous to $A$. A flow $v^\mu$ must have vanishing flux through the timelike boundary.
}
\end{figure}

To illustrate the above conditions, let us consider the examples of asymptotically de Sitter (dS) and anti-de Sitter (AdS) spacetimes. In an asymptotically dS spacetime, with the future and past asymptotic regions cut off by spacelike boundaries, \eqref{Rcond} implies that $R$ must be empty. There are basically three interesting options for $A$ satisfying \eqref{Acond} (see figure \ref{fig:deSitter}): A proper subset of the future cutoff boundary, in which case $\Sigma$ is anchored to that boundary along $\partial A$; all of the future cutoff, in which case $\Sigma$ is any Cauchy slice; or the union of the future cutoff and a proper subset of the past cutoff, in which case $\Sigma$ is pinned to the past cutoff along $\partial A$. On the other hand, in an asymptotically AdS spacetime cut off by spacelike and timelike boundaries, $R$ may be a subset of the timelike boundary. One option is to set $R=\emptyset$ and $A=J^+(\sigma)\cap\partial M$, where $\sigma$ is a Cauchy slice of the timelike boundary. Then $\partial A=\sigma$, and $\Sigma$ is a Cauchy slice anchored to the boundary on $\sigma$. At the other extreme, one could set $R$ equal to the entire timelike boundary and $A$ equal to the spacelike future cutoff. Then $\Sigma$ is again a Cauchy slice, but it is not anchored to any particular boundary Cauchy slice. See figure \ref{fig:AdS}.

We now return to the general discussion. We define a \emph{flow} as a divergenceless future-directed vector field on $M$ with norm $|v^\mu|\ge1$ and vanishing flux through $R$:
\begin{equation}
\nabla_\mu v^\mu = 0\,,\qquad
v^0>0\,,\qquad
-v_\mu v^\mu\ge 1\,,\qquad
\sqrt h\,n_\mu v^\mu|_R = 0\,.
\end{equation}
The reason for imposing a \emph{lower} bound on the norm---as opposed to the upper bound $|v^\mu|\le1$ imposed in the Riemannian case---will become clear shortly. Given a region $A$ and a slice $\Sigma$ homologous to it, the divergenceless condition together with the boundary condition imply
\begin{equation}\label{flux}
\int_A\sqrt h\,n_\mu v^\mu = \int_\Sigma\sqrt h\,n_\mu v^\mu\,.
\end{equation}
The norm condition on $v^\mu$ implies that the flux element $\sqrt hn_\mu v^\mu$ is at least as large as the volume element $\sqrt h$, so
\begin{equation}\label{volume}
\int_\Sigma\sqrt h\,n_\mu v^\mu\ge\int_\Sigma\sqrt h = \vol(\Sigma)\,.
\end{equation}
Combining \eqref{flux} and \eqref{volume}, we have
\begin{equation}\label{lowerbound}
\int_A\sqrt h\,n_\mu v^\mu \ge \vol(\Sigma)\,.
\end{equation}
Thus the flux is \emph{lower}-bounded by the volume of every homologous slice. The one with the largest volume is the ``inverse bottleneck'', giving the strongest lower bound.

The min flow-max cut theorem states that the respective infimum and supremum of the two sides of \eqref{lowerbound} saturate the inequality:
\begin{equation}\label{mfmc3}
\inf_{v^\mu}\int_A\sqrt h\ n_\mu v^\mu = \sup_{\Sigma\sim A}\vol(\Sigma)\,.
\end{equation}
In other words, there are no obstructions other than \eqref{lowerbound} to decreasing the flux of $v^\mu$.

As mentioned above, it may happen that there does not exist a slice homologous to $A$. In that case, since the supremum of the empty set is $-\infty$, the theorem asserts that the flux through $A$ is unbounded below. An example is shown in panel (d) of figure \ref{fig:deSitter}; $A$ has a past-directed normal, so the flux through $A$ of any future-directed timelike vector field is negative. This vector field can have an arbitrarily large norm, so indeed the flux is unbounded below.

\subsection{Proof}\label{sec:proof2}

The proof follows the same steps as the proof of the Riemannian max flow-min cut theorem in subsection \ref{sec:proof}. We will go rather quickly, focusing on the aspects that are different from the Riemannian case.

\paragraph{Step 1:} The future-directed causal (FDC) vector fields $v^\mu$ form a convex set, and within that set, those with $|v^\mu|\ge1$ form a convex subset. The function $-|v^\mu|$ is convex on the set of FDC vectors at a point, but cannot be extended to a (finite) convex function on the full tangent space. We must therefore impose the FDC constraint implicitly. The following convex program has as its solution the left-hand side of \eqref{mfmc3}:
\begin{eqnarray}\label{minflow}
\text{\emph{min flow}}:
\text{minimize }\int_A\sqrt h\ n_\mu v^\mu&\text{ over FDC vector fields }v^\mu\text{ with }\sqrt h\, n_\mu v^\mu|_R=0\,,\nonumber \\ &\text{subject to }\nabla_\mu v^\mu=0\,,\,1-|v^\mu|\le0\,.
\end{eqnarray}

We will give a rough argument that Slater's condition is satisfied. Equation \eqref{Rcond} implies that every point in the interior of $M$ is on a timelike curve that begins and ends on $R^c$. We can therefore cover $M$ with (possibly overlapping) timelike tubes of constant proper thickness $\epsilon$. In each tube we can put a divergenceless FDC vector field with norm greater than 1. Where the tubes overlap, we add the vector fields. The result is a feasible vector field $v^\mu$ with $|v^\mu|>1$ everywhere.

\paragraph{Step 2:} To dualize the \emph{min flow} program \eqref{minflow}, we introduce Lagrange multiplier scalar fields $\psi,\phi$ for the respective explicit constraints, with $\phi\ge0$. The Lagrangian function is
\begin{eqnarray}
L[v^\mu,\psi,\phi] &=&
\int_A\sqrt h\ n_\mu v^\mu+
\int_M\sqrt g\left(-\psi\, \nabla_\mu v^\mu+\phi\, (1-|v^\mu|)\right) \nonumber \\
&=& \int_{\partial M}\sqrt h\ n_\mu v^\mu\, (\chi_A-\psi)+\int_M\sqrt g\left(v^\mu\, \partial_\mu\psi-|v^\mu|\phi+\phi\right).
\end{eqnarray}
Minimizing $L$ with respect to $v^\mu$, keeping in mind the implicit constraints that it is FDC and $\sqrt hn_\mu v^\mu|_R=0$, gives rise to the following dual program:
\begin{multline}\label{maxcut1}
\text{\emph{max cut 1}}:
\text{maximize }\int_M\sqrt g\,\phi\\
\text{ over }\psi,\phi\text{  with }
\psi|_{R^c_0}=\chi_A\,,\,
\psi|_{R^c_-}\ge\chi_A\,,\,
\psi|_{R^c_+}\le\chi_A\,,\,
\partial_\mu\psi\text{ FDC },\,0\le\phi\le|\partial_\mu\psi|\,,
\end{multline}
where $R^c_0$ is the timelike part of $R^c$, $R^c_-$ is the past spacelike part (i.e.\ such that $n_\mu$ is past-directed timelike), and $R^c_+$ is the future spacelike part.

\paragraph{Step 3:} We first eliminate $\phi$ from the \emph{max cut 1} program:
\begin{multline}\label{maxcut2}
\text{\emph{max cut 2}}:
\text{maximize }\int_M\sqrt g\,|\partial_\mu\psi|\\
\text{ over }\psi\text{ with } 
\psi|_{R^c_0}=\chi_A\,,\,
\psi|_{R^c_-}\ge\chi_A\,,\,
\psi|_{R^c_+}\le\chi_A\,,\,
\partial_\mu\psi\text{ FDC }.
\end{multline}

The boundary condition in \emph{max cut 2} looks a bit complicated, but it can be simplified by the following argument. First, we note that the FDC condition on $\psi$ implies that it is non-decreasing along any causal curve. By \eqref{Rcond}, every point in $M$ is on a causal curve starting and ending on $R^c$. The boundary conditions then imply
\begin{equation}\label{psibounds}
0\le\psi\le1\quad\text{everywhere}\,.
\end{equation}
Now suppose that $\psi$ doesn't saturate the inequalities in the boundary condition, for example $\psi<\chi_A$ on some subset $q$ of $R_+^c$. By \eqref{psibounds}, $q$ must be a subset of $A$. Define a new function $\tilde\psi$ that equals 1 in a small neighborhood of $q$ and equals $\psi$ elsewhere. Since $q$ is spacelike, this can be done in such a way that $\partial_\mu\tilde\psi$ is FDC and $|\partial_\mu\tilde\psi|>|\partial_\mu\psi|$. (Effectively, we have added extra level sets to $\psi$ in a neighborhood of $q$.) Thus the objective for $\tilde\psi$ is greater than for $\psi$. So, without changing the supremum, we can require $\psi=\chi_A$ on $R^c_+$. By the same argument, we can require $\psi=\chi_A$ on $R^c_-$. We now have the following program:
\begin{equation}\label{maxcut3}
\text{\emph{max cut 3}}:
\text{maximize }\int_M\sqrt g\,|\partial_\mu\psi|
\text{ over }\psi\text{ with } 
\psi|_{R^c}=\chi_A\,,\,
\partial_\mu\psi\text{ FDC }.
\end{equation}

As in the Riemannian case, we now define the bulk region $r(p)$ ($p\in\R$) by
\begin{equation}\label{rdef2}
r(p):=\{x\in M:\psi(x)\ge p\}\,,
\end{equation}
and the level set $\Sigma(p)$ as the closure of $\partial r(p)\setminus\partial M$ with orientation covector $n_\mu$ parallel to $\partial_\mu\psi$. By the FDC condition on $\partial_\mu\psi$, $\Sigma(p)$ is a slice, and for $0<p<1$ it is homologous to $A$. An important difference to the Riemannian case is that there are no level sets with $p<0$ or $p>1$, which follows from \eqref{psibounds}. We thus have
\begin{equation}
\int_M\sqrt g\,|\partial_\mu\psi| = \int_{-\infty}^{\infty} dp\,\vol(\Sigma(p)) = \int_0^1dp\,\vol(\Sigma(p))\,.
\end{equation}

From here, the argument is identical to the Riemannian case, except with inf replaced by sup: $\int_0^1dp\vol(\Sigma(p))$ is bounded above by the volume of the maximal slice homologous to $A$; conversely, given any slice $\Sigma\sim A$ with corresponding bulk region $r$, the function $\chi_r$ satisfies the constraints of the \emph{max cut 3} program; therefore the solution of \emph{max cut 3} equals the right-hand side of \eqref{mfmc3}.

\subsection{Extensions}\label{sec:extensions}

The extensions of the Riemannian max flow-min cut theorem described in subsections \ref{sec:relation} and \ref{sec:nesting} carry over almost unchanged to the Lorentzian setting. We will state them without proof:
\begin{itemize}
\item If there exist a max cut $\Sigma^*$ and min flow $v^{\mu*}$, then $\Sigma^*$ is spacelike and $v^{\mu*}=n^\mu$ on $\Sigma^*$.
\item If $\Sigma^*$ is unique, then there exists a min flow where $|v^{\mu*}|>1$ everywhere except on $\Sigma^*$.
\item We say that a flow $v^\mu$ can be \emph{diminished} if there exists a vector field $\Delta v^\mu$ with negative flux through $A$ such that $v^\mu+\Delta v^\mu$ is still a flow. A flow that cannot be diminished is a min flow and vice versa. In a non-compact spacetime, we can define a \emph{min flow} as a flow that cannot be diminished, and a globally maximal slice as the locus where all min flows have unit norm. This definition agrees with the globally maximal slice in the presence of a cutoff in the limit that the cutoff is removed.
\item Let $A,B$ be disjoint boundary regions such that there exist unique max cuts $\Sigma(A)$, $\Sigma(AB)$. Then the corresponding bulk regions $r(A)$, $r(AB)$ are nested: $r(A)\subset r(AB)$. (As in the Riemannian case, this can also be proven by an inclusion-exclusion argument.) Also, there exists a flow $v(A,B)$ that simultaneously minimizes the flux through $A$ and $AB$.

\begin{figure}[tbp]
\centering
\includegraphics[width=0.25\textwidth]{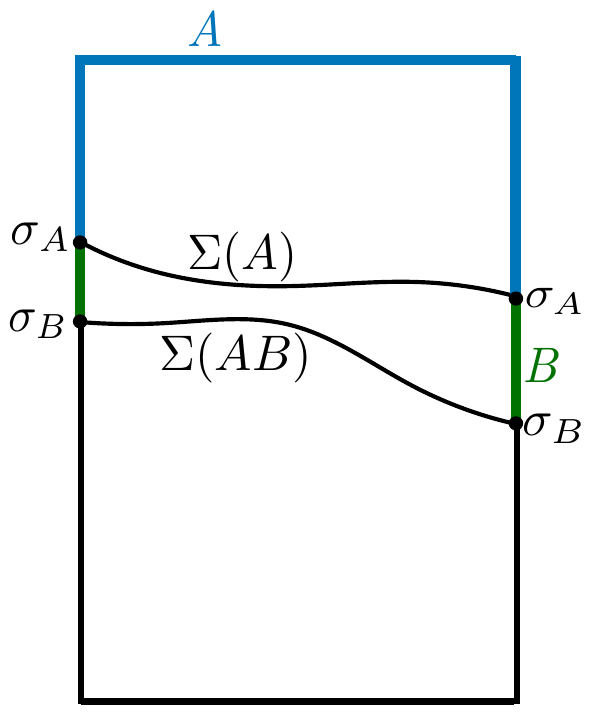}
\caption{\label{fig:nestedmaxcut}
In the asymptotically AdS case, nesting of cuts implies that the maximal slice anchored on the timelike boundary at $\sigma_A$ is entirely to the future of the maximal slice anchored at $\sigma_{AB}$, where $\sigma_A$ is entirely to the future of $\sigma_{AB}$.
}
\end{figure}

\item As a corollary to the previous statement, if $\sigma_A$, $\sigma_{AB}$ are Cauchy slices of the boundary of an asymptotically AdS spacetime such that $\sigma_A$ is entirely to the future of $\sigma_{AB}$, then the maximal-volume slice $\Sigma(A)$ anchored on $\sigma_A$ lies entirely to the future of the maximal-volume slice $\Sigma(AB)$ anchored on $\sigma_{AB}$. See figure \ref{fig:nestedmaxcut}.
\end{itemize}

\subsection{Dilworth's theorem}\label{sec:dilworth}

In this subsection, we show that a special case of the Lorentzian min flow-max cut theorem is a continuum version of Dilworth's theorem from the theory of partially-ordered sets. In a partially-ordered set $P$, two elements $a,b$ are said to be \emph{comparable} if either $a\le b$ or $b\le a$. A \emph{chain} is a subset of $P$ in which any two elements are comparable, while an \emph{antichain} is a subset in which no two elements are comparable. Dilworth's theorem states that the smallest number of chains required to cover $P$ equals the number of elements in the largest antichain.

A causal Lorentzian manifold (i.e.\ one without closed causal curves) is naturally a partially-ordered set with respect to the chronological past, i.e.\ we define $p\le q$ if either $p=q$ or $p\in I^-(q)$. Under this partial order, a timelike curve is a chain and an achronal set is an antichain. Rather than counting discrete chains, we will put a certain measure on a foliation of $M$ by timelike curves. Similarly, rather than counting discrete elements of an anti-chain, we will consider the volume of an achronal set. The continuum Dilworth's theorem will equate the smallest total measure of a foliation to the largest volume of an achronal set.

We assume that $M$ is causal. Set $A=\partial M_+$, the future spacelike part of $\partial M$. Set $R$ equal to the timelike and null parts of $\partial M$. Any slice $\Sigma\sim\partial M_+$ is achronal, as explained below \eqref{relhomdef2}. Conversely, any achronal set $S$ is contained in a slice $\Sigma\sim\partial M_+$.\footnote{
More precisely, $S$ is contained in the limit of slices homologous to $\partial M_+$. Proof: Let $n$ be a bulk region which is a small neighborhood of $\partial M_+$ and set $r:=n\cup J^+(S)$ and $\Sigma$ equal to minus the past boundary of $r$. $\Sigma$ is a slice homologous to $\partial M_+$, and contains $S\setminus n$. In the limit that $n$ is taken vanishingly small, $\Sigma$ contains $S$. (However, in this limit $\Sigma$ may not itself be a slice, since it may partially coincide with $\partial M_+$.)} Therefore
\begin{equation}
\sup_{S\text{ achronal}}\vol(S) = \sup_{\Sigma\sim\partial M_+}\vol(\Sigma)\,.
\end{equation}
Thus the min flow-max cut theorem says
\begin{equation}\label{dilworthcont}
\inf_{v^\mu}\int_{\partial M_+}\sqrt h\,n_\mu v^\mu = \sup_{S\text{ achronal}}\vol(S)\,.
\end{equation}

To argue that \eqref{dilworthcont} is a continuum version of Dilworth's theorem, we will identify a flow with a foliation by future-directed timelike curves equipped with a measure. Henceforth we abbreviate ``future-directed timelike'' as just ``timelike''. The integral curves of a timelike vector field $v^\mu$ are timelike curves that foliate $M$. Furthermore, if $v^\mu$ is divergenceless, then it naturally endows the foliation $F$ with a measure $\mu$, defined by the condition that the total measure of the curves crossing an arbitrary oriented hypersurface $m$ (weighted by the net number of times the curve crosses $m$) equals the flux of $v^\mu$ through $m$. Conversely, a foliation $F$ of $M$ by timelike curves, equipped with a measure, defines a divergenceless timelike vector field. We will say that $F$ covers $M$ densely if, for every slice $\Sigma$,
\begin{equation}\label{densefoliation}
\mu(F\cap\Sigma)\ge\vol(\Sigma)
\end{equation}
(where $F\cap\Sigma$ is the subset of $F$ that intersects $\Sigma$); \eqref{densefoliation} is equivalent to the condition $|v^\mu|\ge1$. Finally, the vector field satisfies $\sqrt h \, n_\mu v^\mu=0$ on $R$ if and only if its integral curves don't begin or end there, which is to say they are inextendible (recall that $R$ is the timelike and null part of $\partial M$). Since every curve then ends on $\partial M_+$, we have
\begin{equation}
\mu(F)=\mu(F\cap\partial M_+) = \int_{\partial M_+}\sqrt h\,n_\mu v^\mu\,,
\end{equation}
and \eqref{dilworthcont} can be written
\begin{equation}\label{dilworthcont2}
\inf_F\mu(F) = \sup_{S\text{ achronal}}\vol(S)\,,
\end{equation}
where the infimum is over dense foliations by inextendible curves. This is clearly a continuum version of Dilworth's theorem.

There is one respect in which \eqref{dilworthcont2} is not strictly analogous to the usual, discrete version of Dilworth's theorem. We have required the timelike curves to be both inextendible and (since they define a foliation) non-intersecting. In the discrete case, neither of these conditions is imposed. It is possible to impose either one, but not both, without changing the result. (If one tries to impose both, it may not be possible to cover the set.) Continuous partially-ordered sets are more flexible in that regard, since the chains (in this case, timelike curves) can be ``squeezed together'' arbitrarily closely without intersecting.

\subsection{Degenerate metric}\label{sec:degenerate}

In certain cases of interest, the Lorentzian manifold $M$ may itself be a timelike submanifold of a larger Lorentzian spacetime, with the metric induced from the larger spacetime. For example, one may be interested in finding the largest-volume\footnote{
For consistency with the rest of this section, we continue to use the term ``volume'', although for submanifolds of codimension two and higher in Lorentzian spacetimes, the term ``area'' is more typically used.} codimension-two spacelike submanifold contained in a given timelike hypersurface. The MFMC theorem equates this volume to the minimal flux of a flow residing on the hypersurface.

One may also be interested in finding the maximal-volume surface within a given \emph{null} submanifold of a Lorentzian spacetime. The induced metric on a null submanifold is degenerate, i.e.\ it has signature $(0,+,+,\ldots)$. Therefore, we describe here the MFMC theorem on a manifold $N$ with such a degenerate metric. (From here on, we work entirely within $N$, without reference to any ambient spacetime it may be embedded in.) Certain modifications to the definitions in subsection \ref{sec:statement2} are needed. First, we allow $N$ to have a null boundary. Second, a \emph{slice} is a spacelike (i.e.\ nowhere null) hypersurface whose normal covector $n_\mu$ is future-directed. (The inverse metric is not defined, so we do not normalize $n_\mu$. A \emph{future-directed} covector is one that contracts positively with any non-zero future-directed null vector.)
Given the non-existence of timelike vectors, we define a flow instead using forms, analogous to calibrations. The metric on $N$ gives rise to a volume form $\omega$, which is the unique $d-1$ form such that, for any slice $\Sigma$, $\vol(\Sigma) = \int_\Sigma\omega$. A \emph{flow} is a closed $d-1$ form of the form $u=\rho \, \omega$ where $\rho$ is a scalar $\ge1$. Since the flux of a flow across the null boundary of $N$ necessarily vanishes, we set $R=\emptyset$ without loss of generality.\footnote{
This reflects the fact that it is impossible to anchor a slice on the null part of $\partial N$. Given a slice $\Sigma$, it can be deformed in a small neighborhood of the null boundary to have an arbitrary intersection with the null boundary at an arbitrarily small cost in volume.\label{foot:nobc}} 
For any region $A$ on $\partial N$,
the analogue of \eqref{mfmc3} is
\begin{equation}\label{mfmc4}
\inf_u\int_Au = \sup_{\Sigma\sim A}\vol(\Sigma)\,.
\end{equation}

\paragraph{Proof:} The left-hand side of \eqref{mfmc4} is the solution of the convex program
\begin{equation}\label{minflow2}
\text{\emph{min flow}}:
\text{minimize }\int_Au\text{ over $u=\rho \, \omega$ }\text{subject to }du=0\,,\,1-\rho\le0\,.
\end{equation}
Let $\tau$ be an arbitrary future-directed 
one-form; $\tau$ will serve as a bookkeeping device, and will drop out in the end. Wedging $\tau$ with $\omega$ 
yields a non-zero top-form with respect to which we can integrate functions on $N$. 
With Lagrange multipliers $\psi,\phi\ge0$ for the respective explicit constraints in \eqref{minflow2}, the Lagrangian function is
\begin{eqnarray}
L[u,\psi]&=&\int_Au +\int_N\left(\phi\, (1-\rho)\, \omega\wedge\tau-\psi \, du\right) \\
&=& \int_{\partial N}\rho\, (\chi_A-\psi)\, \omega
+\int_N\omega\wedge\left(\phi\, \tau-\rho\, \phi\, \tau+\rho \, d\psi\right).
\end{eqnarray}
We now minimize $L$ with respect to $\rho$. On the spacelike part of $\partial N$, we obtain the boundary condition,
\begin{equation}
\psi = \chi_A\,.
\end{equation}
On the null part, since the pullback of $\omega$ vanishes, there is no boundary condition; this implies that there will be no boundary condition on the slice (consistent with the comment in footnote \ref{foot:nobc}). In the bulk, we have
\begin{equation}
\phi\, \omega\wedge\tau = \omega\wedge d\psi\,.
\end{equation}
This fixes $\phi$ in terms of $\psi$. Eliminating $\phi$, the constraint $\phi\ge0$ transfers to $d\psi$, requiring it to be future-directed. 
The dual program is thus
\begin{equation}\label{maxcut4}
\text{\emph{max cut}}:
\text{maximize }\int_N\omega\wedge d\psi
\text{ over }\psi\text{ with } 
\psi|_{\partial N_{\text{spacelike}}}=\chi_A\,,\,
d\psi\text{ future-directed}.
\end{equation}
By the now-standard argument involving level sets, the solution to this program is the right-hand side of \eqref{mfmc4}.

We close the paper with a general comment. In the Introduction, we remarked on the difference between Riemannian and Lorentzian spacetimes, noting that, naively, we might not expect to find an analog of the MFMC theorem beyond the Riemannian case.  We have now seen that in fact the MFMC theorem is not only generalizable to Lorentzian spacetimes, but even further, including to spaces with degenerate metric.  In fact, the power of Lagrangian duality which we have exemplified in a still-limited geometrical context, and which will be utilized further in forthcoming work (e.g.\  \cite{covariantflows,closedsft}), is substantially more far-reaching.  Since vast part of physics can be formulated in terms of  optimization problems, it is tempting to speculate that convex relaxation and Lagrangian duality will become central tools in the subject.

\acknowledgments{
We would like to thank M. Mueller and S. Boyd for very helpful discussions on convex optimization, N. Engelhardt for suggesting a Lorentzian analogue of the max flow-min cut theorem, and S. Aaronson and E. Bachmat for pointing out the connection to Dilworth's theorem. We would also like to thank J. Harper, T. He, and B. Zwiebach for very helpful comments on an earlier draft of the paper. The work of M.H. was supported in part by the National Science Foundation through Career Award No.\ PHY-1053842 and in part by the Simons Foundation through \emph{It from Qubit: Simons Collaboration on Quantum Fields, Gravity, and Information} and through a Simons Fellowshop in Theoretical Physics. V.H.\ was  is supported in part by U.S. Department of Energy grant DE-SC0009999.  M.H. would also like to thank the MIT Center for Theoretical Physics for hospitality.  We would also like to thank the KITP for hospitality during the final stages of this work.
}

\bibliographystyle{JHEP}

\providecommand{\href}[2]{#2}\begingroup\raggedright\endgroup

\end{document}